\documentclass[iop,numberedappendix,apj,]{emulateapj}

\usepackage{epsfig}
\usepackage{amsmath}
\usepackage{rotating}
\usepackage{natbib}
\usepackage{enumerate}
\usepackage{multirow}
\usepackage{array}
\usepackage{appendix}
\usepackage{comment}
\usepackage{color,xcolor}
\usepackage{url}
\usepackage{hyperref}
\hypersetup{colorlinks,linkcolor={blue!50!black},citecolor={blue!50!black},urlcolor={blue!50!black}}
\allowdisplaybreaks[1]
\def\plotonesc#1{\centering \leavevmode
\includegraphics[clip=, width=1.70\columnwidth]{#1}}
\def\plotoneh#1{\centering \leavevmode
\includegraphics[clip=, width=.95\columnwidth]{#1}}

\def\plotoneShrinkMed#1{\centering \leavevmode
\includegraphics[clip=, width=.55\columnwidth]{#1}}

\def\gsim{\;\rlap{\lower 2.5pt
 \hbox{$\sim$}}\raise 1.5pt\hbox{$>$}\;}
\def\lsim{\;\rlap{\lower 2.5pt
   \hbox{$\sim$}}\raise 1.5pt\hbox{$<$}\;}

\newcommand{\Tcmb}{\mbox{$T_{\mbox{\tiny CMB}}$}}
\newcommand{\Tsky}{\mbox{$T_{\mbox{\tiny sky}}$}}
\newcommand{\Tsys}{\mbox{$T_{\mbox{\tiny sys}}$}}
\newcommand{\Trx}{\mbox{$T_{\mbox{\tiny rx}}$}}
\newcommand{\sigRMS}{\mbox{$\sigma_{\mbox{\tiny RMS}}$}}

\begin{document}


\title{Radio Emission from Red-Giant Hot Jupiters}

\author{
Yuka Fujii\altaffilmark{1,2} 
David S. Spiegel\altaffilmark{3,4,5}
Tony Mroczkowski\altaffilmark{6,7}
Jason Nordhaus\altaffilmark{8,9}\\
Neil T. Zimmerman\altaffilmark{10}
Aaron R. Parsons\altaffilmark{11}
Mehrdad Mirbabayi\altaffilmark{5}
Nikku Madhusudhan\altaffilmark{12}
}

\affil{$^1$Earth-Life Science Institute, Tokyo Institute of Technology, 
  Tokyo, 152-8550, JAPAN}
  
\affil{$^2$NASA Goddard Institute for Space Studies, 
  New York, NY 10025, USA}
    
\affil{$^3$Analytics \& Algorithms, Stitch Fix,
  San Francisco, CA 94103, USA}

\affil{$^4$Research \& Development, Sum Labs,
  New York, NY 10001, USA}

\affil{$^5$Astrophysics Department, Institute for Advanced Study,
  Princeton, NJ 08540, USA}

\affil{$^6$National Research Council Fellow}

\affil{$^7$Naval Research Laboratory, 4555 Overlook Ave SW, Washington, DC 20375, USA}

\affil{$^8$Department of Science and Mathematics, National Technical Institute for the Deaf, Rochester Institute of Technology, Rochester, NY 14623, USA}

\affil{$^9$Center for Computational Relativity and Gravitation, Rochester Institute of Technology, Rochester, NY 14623, USA}

\affil{$^{10}$Space Telescope Science Institute, 3700 San Martin Drive, Baltimore, MD 21218, USA}

\affil{$^{11}$Astronomy Department, University of California Berkeley, Berkeley, CA, USA}

\affil{$^{12}$Astronomy Department, University of Cambridge, UK}
\vspace{0.5\baselineskip}

\email{
yuka.fujii@elsi.jp
}

\begin{abstract}
When planet-hosting stars evolve off the main sequence and go through the red-giant branch, the stars become orders of magnitudes more luminous and, at the same time, lose mass at much higher rates than their main-sequence counterparts.
Accordingly, if planetary companions exist around these stars at orbital distances of a few AU, they will be heated up to the level of canonical hot Jupiters and also be subjected to a dense stellar wind.
Given that magnetized planets interacting with stellar winds emit radio waves, such ``Red-Giant Hot Jupiters'' (RGHJs) may also be candidate radio emitters.
We estimate the spectral auroral radio intensity of RGHJs based on the empirical relation with the stellar wind as well as a proposed scaling for planetary magnetic fields. 
RGHJs might be intrinsically as bright as or brighter than canonical hot Jupiters and about 100 times brighter than equivalent objects around main-sequence stars.
We examine the capabilities of low-frequency radio observatories to detect this emission and find that the signal from an RGHJ may be detectable at distances up to a few hundred parsecs with the Square Kilometer Array.
\end{abstract}

  

\section{Introduction}
\label{sec:intro}

Planets with strong magnetic fields may generate radio and/or X-ray emission when interacting with energetic charged particles. 
It is well known that Jupiter emits radio waves from its auroral region due to the cyclotron-maser instability \citep[e.g.][]{wu1979,zarka1998,treumann2006}.
Exoplanets could also generate radio emission through similar mechanisms, depending on their intrinsic magnetic fields and the properties of surrounding plasmas, e.g. stellar wind particles and particles from Io-like moons.
This process could provide an avenue to discover planets that are otherwise extremely difficult to find --- those orbiting highly evolved stars.
Unlike traditional methods of discovering exoplanets, in which the planetary signal is a tiny perturbation on light from the star, low-frequency planetary radio emission might be an arena where planets are vastly brighter than their stars.
This paper is an exploration of the surprisingly diverse range of physical processes that lead to this emission and the prospects for detecting radio emission from planets around giant stars with current and near-future low-frequency radio observatories.

Observations of radio emission from solar system planets imply an empirical relation: the auroral radio power is proportional to the input solar wind energy going into each planetary magnetosphere. This is commonly referred to as the ``radiometric Bode's law'' \citep{desch+kaiser1984,zarka2001}. 
Extrapolating this scaling to exoplanets, radio emission from exoplanetary systems has been examined. 
\citet{farrell1999}, \citet{zarka2001}, and \citet{lazio2004} estimated that a few of the known exoplanetary systems may have radio flux of $\sim$1~mJy level, due to the small orbital distance of the planets.  
\citet{stevens2005} gave improved estimates of the stellar mass-loss rate based on X-ray flux and re-evaluated the radio flux of known exoplanetary systems. 
\citet{griesmeier2005} took account of the high stellar activity in the early stage of the planetary system and proposed that the young system would be a good candidate in which to search for radio emission. 
\citet{griesmeier2007a, griesmeier2007b} discussed the effects of the detailed properties of stellar wind in the proximity of the stars. They considered not only the kinetic energy of the stellar wind but also the magnetic energy of the stellar wind and coronal mass ejection. 
Note that in these papers, the scaling relations for the planetary magnetic field differ from paper to paper.
In \citet{reiners2010}, the authors adopted a new scaling relation of planetary magnetic fields based on \citet{christensen_et_al2009}.
\citet{jardine2008} modeled the reconnection between the magnetic field of a close-in planet and that of the host star to obtain estimates that are not based on the ``radiometric Bode's law.'' 
Although observational searches for these radio signatures are underway, no clear detection has been claimed \citep{bastian2000,george2007,smith2009,stroe2012,hallinan2013,murphy2015}, while there are some promising initial results \citep{lecavelier_et_al2013,sirothia2014}.

When stars less than $\sim$8~$M_\sun$ evolve off the main sequence, they evolve through the red-giant branch (RGB) and the asymptotic-giant branch (AGB) phases where their radii and luminosities increase by orders of magnitude.
Jovian planets in orbit around such stars can migrate inward or outward due to the interplay between tidal torques and the mass-loss process on the post-main sequence \citep{nordhaus_et_al2010, kunitomo2011, mustill+villaver2012, spiegel2012, nordhaus+spiegel2013}.
During this time, such planets can be transiently heated to hot-Jupiter temperatures ($\gtrsim$1000~K) at distances out to tens of AU, depending on the star's mass; such planets are termed ``Red-Giant Hot Jupiters'' (RGHJs), though the term can refer to planets orbiting either RGB and AGB stars \citep{spiegel+madhusudhan2012}.
Such planets are also subject to interactions with a massive (but slow) stellar wind, as the mass-loss rate of evolved stars is significant, ranging from $\sim$10$^{-8}$~$M_\odot$~yr$^{-1}$ to $\sim$10$^{-5}$~$M_\odot$~yr$^{-1}$ with the highest values for AGB stars \citep[e.g.,][]{reimers1975, schild1989, vassiliadis1993, schoier2001, vanloon2005}. 
On the assumption that the radio emission is correlated with the stellar wind, planetary companions around evolved stars could also generate bright radio emission. 

Based on this speculation, \citet{ignace2010} examined radio emission from known substellar-mass companions around cool evolved stars.
They found that the low ionization fraction of the stellar wind of evolved stars suppresses their radio emission and leads to weak radio emission. 
In addition, they considered a scenario in which post-bow shock heating could produce ionized hydrogen atoms. This indicates a configuration of radio-wave generation very different from that of the solar system, and the lower-limit estimate, at least, is well below the detection limit.

In this paper, we consider a further plausible scenario, where the accretion of the massive stellar wind onto the planet would emit UV and X-ray photons that ionize the stellar wind in the vicinity of the planet.
The ionized stellar wind particles would then interact with the planetary magnetic field in the same way as the solar system planets do.
Thus, by extrapolating the ``radiometric Bode's law,'' we provide more optimistic estimates compared to the previous result. 
In addition, we introduce two major advances beyond the work of \citet{ignace2010} that are related to the observability of radio emission from RGHJs.
First, we estimate the plasma frequency cut-off of the stellar wind, which turns out to be one of the major obstacles to detecting radio emission from RGHJs.
Second, we discuss the planetary parameter space to search for radio emission, employing scaling laws for the planetary magnetic field, noting that the survey of exoplanets around highly evolved stars is not complete.

In Section \ref{s:assumptions}, we present the framework to obtain the frequency and the flux of planetary auroral radio emission, and we describe our models for the stellar wind and planetary magnetosphere.
Section \ref{s:result} presents the estimates of the spectral radio flux of RGHJs and compares the predictions with what might be expected from canonical hot Jupiters as well as those from Jupiter-twins.
Section \ref{s:observability} gives the prospects for the signal detection with the current/future instruments. 
Estimates for the known late-type (M-type) evolved stars are also included. 
Finally, Section \ref{s:conc} concludes the paper with a brief summary.

\section{Model}
\label{s:assumptions}

In this section, we describe our framework to estimate radio emission from RGHJs. 
First, we introduce our scheme to compute the frequency and intensity of planetary radio emission in Sections \ref{ss:model_frequency} and \ref{ss:model_intensity}, respectively. 
Then, two ingredients for the framework --- the strength of the planetary magnetic field and the properties of the stellar wind --- will be presented in Sections \ref{ss:magneticfield} and \ref{ss:stellarwind}, respectively. 
In Section \ref{ss:ionization}, we consider the ionization around the planets, which is a crucial factor to determine the efficiency of the interaction between the planetary magnetosphere and the stellar wind.

\subsection{Frequency of Radio Emission}
\label{ss:model_frequency}

As ionized electrons flow along planetary magnetic field lines, auroral radio waves are emitted at the local cyclotron frequency.
The upper limit is around the cyclotron frequency of the planetary surface magnetic field, $\nu_{\rm cyc,\,max}$: 
\begin{equation}
\nu_{\rm cyc,\,max} = \frac{eB}{2\pi m_e c} \approx 28 {\rm~MHz} \left( \frac{B}{10 \rm~G} \right) \label{eq:fcyc}
\end{equation}
where $B$ is the strength of the magnetic field at the planetary surface, $e$ and $m_e$ are the charge and mass of electrons, respectively, and $c$ is the speed of light. 

Radio emission from an exoplanet is observable from the ground (on Earth) only when its frequency is greater than both the plasma frequency of Earth's ionosphere $\nu_{\rm plasma}^\oplus$ and the maximum plasma frequency along the line of sight $\nu_{\rm plasma}^{\rm los}$: 
\begin{equation}
\nu_{\rm cyc,\,max} > \nu_{\rm plasma}^\oplus \;\;\; \mbox{and} \;\;\; \nu_{\rm cyc,\,max} > \nu_{\rm plasma}^{\rm los}
\end{equation}
The plasma frequency may be expressed as
\begin{eqnarray}
\nu_{\rm plasma} & = & \sqrt{\frac{n_e e^2}{\pi m_e}} \\
 & = & 8979 {\rm~Hz} \times \left( \frac{n_e}{\rm cm^{-3}} \right)^{1/2} \, .
\label{eq:fplasma}
\end{eqnarray}
In the Earth's ionosphere, the electron number density is less than $10^6$~cm$^{-3}$, which implies that $\nu_{\rm plasma}^\oplus \lsim 10$~MHz. 
Along the line of sight, the maximum plasma frequency $\nu_{\rm plasma}^{\rm los}$ is typically governed by the density of stellar wind particles around the planet, which will be specified in Section \ref{ss:stellarwind} below.

Another factor that affects the radio emission is the plasma frequency at the site of radio-wave generation. This is because the cyclotron-maser instability as a mechanism to generate intense radio emission is efficient when the {\it in situ} plasma frequency is small in comparison to the local cyclotron frequency \citep[][]{treumann2006}.
Future work will be required to estimate the local plasma density and examine this condition.
In this paper, we assume that this condition is satisfied at least at some region along the poloidal magnetic field lines.

\subsection{Flux of Radio Emission}
\label{ss:model_intensity}

The auroral radio spectral flux of exoplanets observed at the Earth, $F_{\nu}$, can be expressed by:
\begin{equation}
F_{\nu} = \frac{P_{\rm radio}}{\Omega l^2 \Delta \nu}
\label{eq:Fnu}
\end{equation}
where $P_{\rm radio}$ is the energy that is deposited as auroral radio emission of the considered frequency range, $\Omega$ is the solid angle of the emission, $l$ is the distance between the target and the Earth, and $\Delta \nu$ is the frequency bandwidth. 

We estimate the radio emission of exoplanets, $P_{\rm radio}$, simply by scaling the Jovian auroral radio emission, $P_{\rm radio,\,J}$, with the input energy from stellar wind, in the same manner as \citet{griesmeier2005,griesmeier2007a,griesmeier2007b}.
The scaling is based on the empirical/apparently good correlation between the radio emission intensity of solar system planets and the input kinetic energy, $P_{\rm inp,\,k}$, or the magnetic energy of the solar wind, $P_{\rm inp,\,m}$, \citep[the ``radiometric Bode's law'';][]{desch+kaiser1984,zarka2001}, i.e.,
\begin{eqnarray}
P_{\rm radio} &\propto & P_{\rm inp} \\
P_{\rm inp,\,k} &=& m_p n v^3 \cdot \pi r_{\rm mag} ^2 \label{eq:Pinp_kin}, \\
P_{\rm inp,\,m} &=& (B_{\star\bot }^2/8 \pi  ) \, v \cdot \pi r_{\rm mag} ^2 \label{eq:Pinp_mag} \, ,
\end{eqnarray}
where $m_p$ is the proton mass, $n$ is the number density of the stellar wind, $v$ is the relative velocity of the stellar wind particles to the planet, $ B_{\star\bot }$ is the interstellar magnetic field perpendicular to the stellar wind flow, and $r_{\rm mag}$ is the distance from the center of the planet to the magnetic stand-off point, described below.
It is not clear from observations of the solar system planets which of the above two relations (Eqs.~\ref{eq:Pinp_kin} and \ref{eq:Pinp_mag}) most accurately captures the ``true'' relationship between input ingredients and output radio power \citep{zarka2001}.
In this paper, we assume that the radio emission scales with the input kinetic energy (i.e., that $P_{\rm radio} \propto P_{\rm k,\,inp}$, as per Eq.~\ref{eq:Pinp_kin}) and consider the possible effects of a dense stellar wind.

Note that if Eq.~\ref{eq:Pinp_mag} is actually the better predictor of radio power, it is difficult to estimate the emission from RGHJs at this point,  because magnetic fields of evolved stars are not well constrained for highly evolved stars; most observations set only upper limits \citep[e.g.,][]{konstantinova2010,konstantinova2013,petit2013,tsvetkova2013,auriere2015}. 
In the case of the M-type giant EK boo, a surface magnetic field $\sim$0.1-10~G has been measured.
In this particular case, given the large stellar radius ($R_\star \sim 210 R_{\odot }$), the magnetic moment may be about $10^3$$\times$ larger than that of the Sun and therefore may also increase the planetary radio emission.
However, due to our present ignorance about the strength of magnetic fields of evolved stars, we leave the magnetic model of radio emission for RGHJs for future work.

In reality, the total power of Jovian auroral emission varies greatly over time.
The average is of the order of $1.3\times 10^{10}$~W. During highly active periods, it averages more like $8.2\times 10^{10}~$W, and the emission can reach powers as high as $4.5 \times 10^{11}$~W during peak activity \citep{zarka_et_al2004}.
Here, we employ $P_{\rm radio,\,J}=2.1\times 10^{11}$~W as a canonical value, following \citet{griesmeier2005,griesmeier2007b}. 

The magnetic stand-off point is where stellar wind ram pressure and planetary magnetic pressure are in approximate balance:
\begin{equation}
m_p n v ^2 \sim \frac{B^2}{8\pi}\left( \frac{r_{\rm mag}}{R_p} \right)^{-6}  \label{eq:stand-off}. 
\end{equation}
The outflow ram pressures from heated planets are negligible compared to these pressures, as discussed in Appendix \ref{ap:outflow}. 
Therefore,
\begin{equation}
r_{\rm mag} \sim R_p \left( 8\pi m_p n \right) ^{-1/6} v^{-1/3} B^{1/3} \label{eq:stand-off-radius} 
\end{equation}
The radius obtained for Jupiter from this equality is about half of the actual magnetospheric radius \citep[][]{griesmeier2005}. 
To estimate $r_{\rm mag}$ for RGHJs, we scale this radius according to the parameter dependence of equation~(\ref{eq:stand-off-radius}). 

For exoplanets, we assume that the solid angle of the emission ($\Omega$) is the same as that of Jupiter.
In reality, the solid angles of auroral radio emission from  Jupiter, Saturn, and Earth are $\sim$1.6, $\sim$6.3, and $\sim$3.5 steradians, respectively \citep{desch+kaiser1984}, which are on the same order and will not significantly affect our order-of-magnitude estimate of radio emission. 
The bandwidth, $\Delta \nu$, is assumed to be proportional to the representative frequency of the emission, which is the cyclotron frequency, following \citet{griesmeier2007b}.

\subsection{Assumptions for Planetary Magnetic Field}
\label{ss:magneticfield}

In order to obtain the frequency and the intensity of the radio emission, we need to compute the strength of the magnetic field at the planetary surface, $B$. 
We do so simply by scaling the Jovian magnetosphere at the surface $B_J \sim 10$ G according to the planetary mass and age, based on the scaling relation described below.

Several scaling relations for planetary magnetic field strength have been proposed \citep[e.g.][]{busse1976,russel1978,stevenson1979,mizutani1992,sano1993,starchenko2002,christensen2006,christensen_et_al2009}, which are summarized and compared with numerical simulations in \citet{christensen2010}.
We employ the scaling law proposed in \citet{christensen_et_al2009} and used in \citet{reiners2010} to explore the evolution of planetary magnetic fields:
\begin{equation}
B_{\rm dynamo} ^2 \propto f_{\rm ohm}\;\rho_{\rm dynamo}^{1/3} \;  (Fq_o)^{2/3} \, , \label{eq:Bscaling} 
\end{equation}
where $B_{\rm dynamo}$ is the mean magnetic field in the dynamo region, $f_{\rm ohm}$ is the ratio of ohmic dissipation to the total dissipation, $\rho_{\rm dynamo} $ is the mean density in the dynamo region, $F$ is an efficiency factor of order unity, and $q_o$ is the convected flux at the outer boundary of the dynamo region (see \citealt{christensen_et_al2009} for a comprehensive description).
This scaling law is based on the assumption that the ohmic dissipation energy is a fraction of the available convected energy and was found to be in good agreement with both the numerical experiments (over a wide parameter space) and with known objects from the Earth to stars. 
Here, $f_{\rm ohm}$ and $F$ are assumed to be constant for the bodies considered in this paper. 
Dipole magnetic field strength at the planetary surface, denoted by $B$, is then scaled by
\begin{equation}
B \propto B_{\rm dynamo} \left( \frac{r_{\rm dynamo}}{R_p} \right)^3 \, .  \label{eq:Bscaling2}
\end{equation}
where $r_{\rm dynamo}$ is the radius of the outer boundary of the dynamo region. 

The scaling law for $B_{\rm dynamo}$, Equation (\ref{eq:Bscaling}), is reasonable only for rapidly rotating objects.
Unlike canonical hot Jupiters, RGHJs are not tidally locked to their host stars, so they probably have the rapid rotation needed for Eq.~\ref{eq:Bscaling} to be a useful ansatz \citep{spiegel+madhusudhan2012}.
In this paper, we assume that RGHJs indeed are rapidly rotating so that they generate planetary magnetic fields through the same mechanism as Jupiter.

In order to evaluate $\rho _{\rm dynamo}$ and $q_o$, we need a model of the internal planetary structure. 
We consider Jupiter-like gaseous planets and assume that the planetary radius is constant at $R_p = R_{p,{\rm J}}$, as numerical calculations show that the radii of gaseous planets over the range of $0.1 M_{p,{\rm J}} < M_p < 10M_{p, {\rm J}}$ (with core mass less than 10\%) converge to $0.8 R_{p,{\rm J}} < R_p < 1.2R_{p, {\rm J}}$ within 1 Gyr \citep{fortney2007, spiegel+burrows2012, spiegel+burrows2013}.
For the density profile, we follow \citet{griesmeier2007b} and  assume a polytropic gas sphere with index $n=1$, which results in:
\begin{equation}
\rho [r] = \left( \frac{\pi M_p}{4 R_p^3} \right) \frac{\sin \left[ \pi \frac{r}{R_p} \right]}{\left( \pi \frac{r}{R_p} \right)} \, . \label{eq:rho_r}
\end{equation}
We determine the radius of the outer boundary of the dynamo region, $r_{\rm dynamo}$, by assuming that the hydrogen becomes metallic when $\rho (r)$ exceeds the critical density $\rho_{\rm crit} = 0.7\,\mbox{g/cm}^3$ \citep{2006explanets_guillot, griesmeier2007b}.
The density of the metallic core, $\rho _{\rm dynamo}$ is obtained by averaging the density in the core. 
In the case of Jupiter, $r_{\rm dynamo, J} = 0.85 R_{\rm J}$ and $\rho_{\rm dynamo, J} = 1.899~{\rm g/cm}^3$.

The scaling of convected heat flux at the outer boundary, $q_o$, is obtained by dividing the age-dependent net planetary luminosity, $L_p$, by the surface area of the core region, i.e., $4\pi r_{\rm dynamo}^2$. 
The time-dependent luminosity is taken from equation (1) of \citet{burrows_et_al2001} \citep[see also][]{marley2007}. 
Ignoring the relatively weak dependence on average atmospheric Rosseland mean opacity leads to:
\begin{equation}	
L_p \sim 6.3\times10^{23} \; {\rm erg} \; \left( \frac{t}{4.5 \rm~Gyr} \right)^{-1.3} \left( \frac{M_p}{M_{\rm J}} \right)^{2.64} \, .
\label{eq:burrowsLum}
\end{equation}
Therefore, we have
\begin{equation}
q_o \sim q_{o, {\rm J}} \left( \frac{t}{4.5 \rm~Gyr} \right)^{-1.3} \left( \frac{M_p}{M_{\rm J}} \right)^{2.64} \left( \frac{r_{\rm dynamo}}{r_{\rm dynamo,\,J}} \right)^{-2}\, .
\label{eq:burrowsHeatFlux}
\end{equation}

\subsection{Assumptions for Stellar Wind}
\label{ss:stellarwind}

The other key ingredient for radio emission is the stellar wind.  
The number density of particles in the stellar wind, $n$, can be expressed as
\begin{equation}
n = \frac{\dot M_\star}{4\pi a^2 m_p v_{\rm sw}} \, ,
\label{eq:n}
\end{equation}
where $\dot M_\star$ is the stellar mass-loss rate, $a$ is the orbital distance from the star, $m_p$ is the proton mass, and $v_{\rm sw}$ is the velocity of the stellar wind.
For the solar wind, $\dot M_\odot \sim 2\times 10^{-14} M_{\odot}$/yr and $v_{\rm sw} \sim 400~{\rm km/s}$ \citep[e.g.,][]{hundhausen1997}.

The mass-loss rate of red giants is typically $\dot M_\star \sim 10^{-8}-10^{-7} M_{\odot}$/yr \citep{reimers1975}, and the rate can be as high as $10^{-5} M_\odot{\rm /yr}$ during the AGB phase \citep{schild1989, vassiliadis1993, schoier2001, vanloon2005}.
Therefore, we have
\begin{equation}
\frac{\dot M_\star}{\dot M_{\odot}} \sim 10^6 - 10^9 \, . \label{eq:scale_Mdot}
\end{equation}

The stellar wind velocity, $v_{\rm sw}$, becomes smaller as the star evolves.
The wind velocity is typically of the order of the escape velocity at a distance of several times the stellar radius \citep{suzuki2007}, i.e., $\sim$$\sqrt{2GM_{\star}/R'_{\star}}$, where $R'_{\star}$ is several times $ R_{\star}$.
For a star with radius $R_{\star }=100~R_{\odot }$, this results in $v_{\rm sw}\sim 30~{\rm km/s}$.
Therefore, a solar-mass red giant with radius $R_{\star}=100R_{\odot}$ produces a stellar wind that is slower by an order of magnitude than that of the Sun, $v_{\odot}$. 

Based on equation (\ref{eq:scale_Mdot}), the number density of the stellar wind (equation \ref{eq:n}) is normalized as follows:
\begin{eqnarray}
n &=& 1.8 \times 10^6 \; {\rm cm^{-3}} \times \left( \frac{a}{5 \; {\rm AU}} \right)^{-2} \notag \\
&&\times \left( \frac{\dot M_\star}{10^{-8} M_{\odot }{\rm /yr}} \right) \left( \frac{v_{\rm sw}}{30~\mbox{km/s}} \right)^{-1} \, . \label{eq:n_normalized}
\end{eqnarray}

The velocity term in equation (\ref{eq:Pinp_kin}), which is the relative velocity between the planet and the stellar wind particles falling onto the planet, depends on the stellar wind velocity, the infall velocity (i.e., the acceleration due to planet's gravitational field), and the planetary orbital velocity.
All three of these terms might be of the same order of magnitude:
as described above, $v_{\rm sw}$ is $\sim$30~km/s;
the infall velocity from the planetary gravity is $\sim$10-25~km/s depending on the planetary mass (ranging from $1 M_{\rm J} to 10 M_{\rm J}$);
and the orbital velocity is $10-30~{\rm km}\,{\rm s}^{-1}$ depending on the orbital distance (ranging from $1-5~{\rm AU}$).
In the next section, we simply consider
\begin{equation}
\frac{v}{v_{\odot}} \sim 10^{-1} \, \label{eq:scale_v}
\end{equation}
a fiducial value for the normalization.

\subsection{Ionization of Stellar Wind Particles Around the Planet}
\label{ss:ionization}

As discussed in \citet{ignace2010}, as stars evolve, the ionization fraction of the stellar wind diminishes to the order of $\sim$10$^{-3}$ \citep{drake1987}.
Since only charged particles interact with a planetary magnetic field, this suggests inefficient interaction with the planetary magnetosphere and hence,  low input energy for radio emission.
However, for a highly evolved star, the velocity of the stellar wind eventually becomes slower than the escape velocity of the planetary companion and hence, stellar wind particles will accrete onto the planets.
\citet{spiegel+madhusudhan2012} considered Bondi--Hoyle accretion where the accretion radius is
\begin{eqnarray}
R_{\rm acc} &=& \frac{2GM_p}{v_{\rm rel}^2} \\
&\sim & 4 ~R_{\rm J} \left( \frac{M_p}{M_{\rm J}} \right) \left( \frac{v_{\rm rel}}{30 ~{\rm km~s^{-1}}} \right)^{-2} \, .
\end{eqnarray}
The accretion luminosity $L_{\rm acc}$ and temperature $T_{\rm acc}$ are
\begin{eqnarray}
L_{\rm acc} \sim &&  10^{25} \; {\rm erg\;s}^{-1} \left( \frac{\dot M_\star}{10^{-8} M_{\odot }/{\rm yr}} \right)  \notag \\
&& \times \left( \frac{M_p}{M_{\rm J}} \right)^3 \left( \frac{M_\star}{M_{\odot }} \right)^{-2} \, . \label{eq:Lacc}\\
T_{\rm acc} \sim && 2 \times 10^5 ~\mbox{K} \left( \frac{M_p}{M_{\rm J}} \right) \, . \label{eq:Tacc}
\end{eqnarray}
The accretion onto planets, therefore, leads to emission of UV/X-ray photons whose characteristic energy $\sim$$k_B T_{\rm acc}$ exceeds the ionization energy of hydrogen, $E_{\rm Rydberg} = 13.6$~eV ($\lambda = 91.2\rm~nm$).
Therefore, UV/X-ray radiation from accretion will create a local ionized region around the planet. 

Let us consider the ionization profile around the planet. 
We suppose a state where the ionization and recombination rates are in equilibrium. 
Denoting the ionization fraction by $x$, the equilibrium state at a distance $r$ from the planet may be represented by 
\begin{equation}
\frac{\dot N_X}{ 4 \pi r^2 } e^{- \tau } n ( 1-x ) \sigma _{\rm H} [E_{\rm photon}] = (n x )^2 \beta [T_e] \label{eq:equilibrium} 
\end{equation}
\begin{equation}
\tau = \int _{R_{\rm J}}^r n(1-x) \sigma_{\rm H} [E_{\rm photon}] dr \, ,
\end{equation}
where $\dot N_X$ is the source rate of the photons that can ionize hydrogen, $E_{\rm photon}$ is the energy per photon, and $\sigma _{\rm H}$ is the cross-section of H atoms for X-ray photons.
Per \citet{verner1996}, $\sigma _{\rm H}$ scales as
\begin{equation}
\sigma _{\rm H} [E_{\rm photon}] \sim 6.3 \times 10^{-18} \;{\rm cm}^2 \cdot \left( \frac{E_{\rm photon}}{E_{\rm Rydberg}} \right)^{-3} \, , \label{eq:sigma_H}
\end{equation}
where $\beta [T_e]$ is the ``class B'' recombination coefficient as a function of the electron temperature, $T_e$.
We adopt the value at $T_e \sim 10^4$~K, $\beta \sim 2.6 \times 10^{-13}\;{\rm cm^3/sec} $, in the following \citep{pequignot1991}; when the electron temperature is varied from $10^3$ to $10^5$~K, $\beta $ varies $1.5 \times 10^{-12}-3.2 \times 10^{-14}\;{\rm cm^3}\,{\rm sec}^{-1}$.

The source rate is obtained by counting the number of photons with energy exceeding $E_{\rm Rydberg}$, which is approximately given by dividing the X-ray accretion luminosity by the characteristic photon energy produced:
\begin{equation}
\dot{N}_X \sim \frac{L_{\rm acc}}{k_{\rm B} T_{\rm acc}} \, ,
\end{equation}
where $k_{\rm B}$ is Boltzmann's constant.
As a result, 
\begin{eqnarray}
\dot{N}_X \sim  
  \left\{
    \begin{array}{ll}
      4 \times 10^{35} \;  {\rm sec^{-1}} & \;\;\;(M_p = M_{\rm J}) \\
      4 \times 10^{37} \; {\rm sec^{-1}} & \;\;\;(M_p = 10 M_{\rm J})
    \end{array} \, .
  \right.
\end{eqnarray}

For accretion onto very massive planets, the characteristic photon energy is so high that the released energetic electrons may also ionize other atoms in the vicinity.
The cross-section\footnote{This is close to the geometric cross-section of the Bohr radius.} of hydrogen for electrons is $\sim$4$\times$10$^{-17}$~cm$^2$ \citep{fite1958}, which implies a mean free path for ionized electrons of $\sim$2$R_{\rm J}$ in the surrounding medium.
As a result, nearly 2/3 of released energetic electrons will ionize a hydrogen atom within $\sim$2$R_{\rm J}$, and more than 99\% will ionize a hydrogen atom within $\sim$10$R_{\rm J}$.

In principle, photons with energy $E_{\rm photon}$ have the potential to ionize $E_{\rm photon}/E_{\rm Rydberg}$ hydrogen atoms.
To account for this, we consider two limiting possibilities:
After an ionizing collision, the energy can (\emph{i}) be split evenly between the two electrons, or it can (\emph{ii}) go entirely into the kinetic energy of one electron and not at all into that of the other (of course, any split in between these extremes is possible, too).
Note that conservation of momentum and energy imply that the proton will not acquire a significant fraction of the energy of the collision.\footnote{Were it otherwise, we would have to take into account what fraction of a rubber ball's kinetic energy is imparted to the kinetic energy of the Earth when bouncing a ball.}  

Scenario (\emph{i}) implies a cascade where an electron with energy $E_{\rm in}$ ionizes an atom, producing two electrons (an ionizing electron plus the released electron) with energy $ ( E_{\rm in} - E_{\rm Rydberg} ) /2 $ for each.
For example, in an idealized case with $M_p = 10 M_{\rm J}$, $k_B T_{\rm acc} \sim 172\,{\rm eV} \sim 12 \, E_{\rm Rydberg}$, a photoionization could produce an electron with energy of $(12-1) = 11 E_{\rm Rydberg} $, then a second ionization by that electron would produce two photons with energy of $(11-1)/2 = 5 E_{\rm Rydberg}$, and the third ionization by these two photons would produce four photons with energy of $(5-1)/2 = 2 E_{\rm Rydberg}$, etc.
The cascade can proceed to the fourth order at the maximum.

Under scenario (\emph{ii}), the cascade proceeds with the initial $E_{\rm in}$ electron leading to an electron with kinetic energy $E_{\rm in} - E_{\rm Rydberg}$ and another electron with zero kinetic energy.
Clearly, this cascade can produce a maximum total of $E_{\rm in} / E_{\rm Rydberg}$ free electrons.

In reality, not all released electrons result in further ionization interactions.
If $\xi$ represents the fraction of released electrons that proceed to the next ionization, then the number of ionized atoms $N_i$ released through this cascade (\emph{i}) is
\begin{eqnarray}
  \nonumber N_i & = & (1-\xi) + 2\xi (1-\xi) + 4 \xi^2 (1-\xi) + 8 \xi^3 \\
  \label{eq:N_i1} & = & 1 + \xi + 2 \xi^2 + 4 \xi ^3 \, .
\end{eqnarray}
Alternatively, cascade (\emph{ii}) leads to
\begin{eqnarray}
  \nonumber N_i & = & \frac{1 - \xi^k}{1 - \xi} \\
  \label{eq:N_i2}  & = & 1 + \xi + \xi^2 + \cdots + \xi^{k - 1} \, ,
\end{eqnarray}
where $k \equiv E_{\rm in} / E_{\rm Rydberg}$ is the maximum number of ionizations for the given initial electron energy.
This limit leads to a value for $N_i$ that is not dramatically different from that of limit (\emph{ii}).
In Appendix \ref{sec:AppendixA}, we estimate the M{\o}ller scattering cross-section and show that cascade (\emph{ii}) --- unequal recoil energies --- is probably more realistic.

Ultimately, $\dot{N}_X $ in equation (\ref{eq:equilibrium}) is replaced by 
\begin{equation}
\dot{N}_X \rightarrow N_i \dot{N}_X \, .
\end{equation}
For a $10 M_{\rm J}$ planet, $N_i$ is probably approximately in the range 5---10.

\begin{figure}[htbp]
   \plotoneh{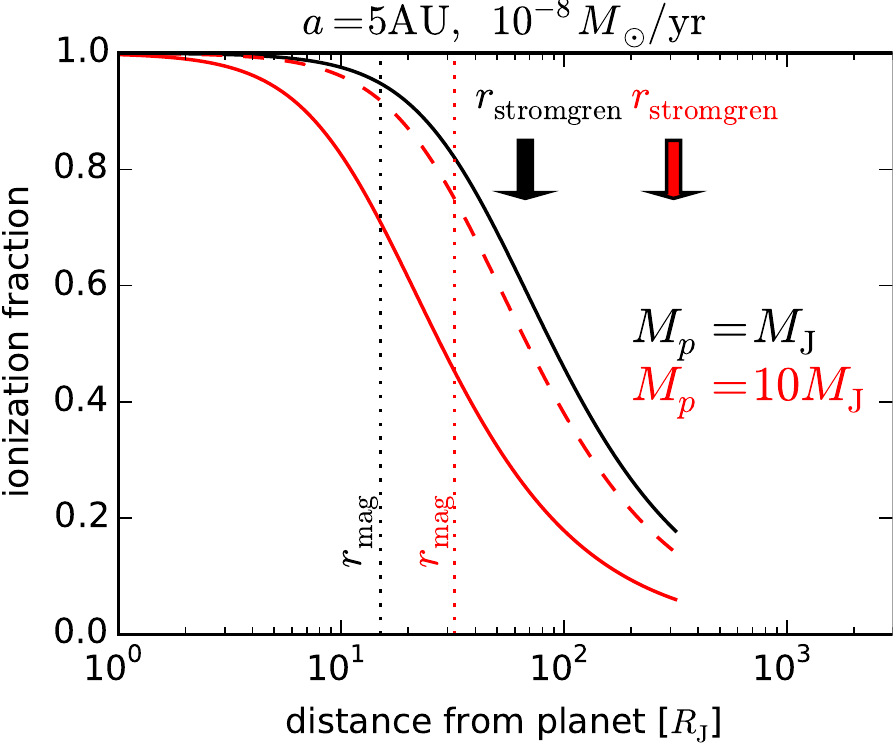}
   \caption{Profile of ionization fraction measured from the surface of the planet, due to UV/X-ray from the accretion of stellar wind onto the planet.
     Solid lines show the solutions without the correction for the secondary ionization by ionized electrons, and the dashed line shows the solution for a  10$M_{\rm J}$ planet taking the correction into account with efficiency factor $N_i=6$.
     The vertical arrows show the Str\"omgren radius estimated simply using  equation (\ref{eq:stromgren}).
     The dotted vertical lines indicate the location of the magnetic stand-off radii, $r_{\rm mag}$, obtained using equation (\ref{eq:stand-off-radius_RGHJ}) below. }
  \label{fig:ionizationfraction}
\end{figure}

The ionization fraction $x$ as a solution of equation (\ref{eq:equilibrium}) is shown in Figure \ref{fig:ionizationfraction}. 
When $\tau \sim n \sigma_H r$ is much smaller than unity and thus the $e^{-\tau}$ term can be ignored (as is the case for $M_p = 10 M_{\rm J}$), the solution is simply
\begin{eqnarray}
x[r] &=& \frac{-1 + \sqrt{1+4C[r]}}{2C[r]} \\
C[r] &\equiv &   \frac{4 \pi n \beta [T_e] r^2}{\dot{N} \sigma_H[E_{\rm photon}]} \, .
\end{eqnarray}
The solid lines show the ionization fraction corresponding to no additional ionization by electrons (i.e., $\xi=0$, or $N_i = 1$) and the dashed line shows the solution with $N_i=6$ for a $10 M_{\rm J}$ planet.
In the figure, the vertical lines show the magnetic stand-off radius obtained using equation (\ref{eq:stand-off-radius_RGHJ}) below.
While the photon rate is larger for more massive planets, the strong dependence of the cross-section on photon energy (equation \ref{eq:sigma_H}) leads to a decrease in the radius of the ionized region.
Nevertheless, {\it a substantial amount of ionized plasma is expected around the magnetic stand-off radius, despite the initially low ionized fraction of the stellar wind}.

The extent of the ionized region may also be roughly estimated as the Str\"omgren radius \citep{stromgren1939}: 
\begin{equation}
r_{\rm stromgren} = \left( \frac{3}{4\pi} \frac{\dot N_X}{n^2 \beta } \right)^{1/3} \, ,
\end{equation}
which gives
\begin{eqnarray}
r_{\rm stromgren} \sim  
  \left\{
    \begin{array}{ll}
      67 ~R_J & \;\;\;(M_p = M_{\rm J}) \\
      310 ~R_J & \;\;\;(M_p = 10 M_{\rm J})
    \end{array}
  \right. \label{eq:stromgren}
\end{eqnarray}
for a planet 5~AU from a red giant host. 
The Str\"omgren radii for 1-$M_{\rm J} $ and 10-$M_{\rm J}$ planets are also indicated in Figure \ref{fig:ionizationfraction}.

Note that this is a very interesting and different regime of the Str\"omgren sphere from the commonly-considered case (photoionization around O/B-type stars).  Around RGHJs, the Str\"omgren sphere does not delineate a sharp edge of ionization, because of the smaller source rate and smaller photon-hydrogen cross-section (in this case parameter ``$a$'' in equation 13 of \citealt{stromgren1939} is not small) on account of X-ray photons interacting more weakly with neutral hydrogen than UV photons near the ionization limit.
The Str\"omgren radii are indicated by vertical arrows in Figure \ref{fig:ionizationfraction}. 
Crucially, X-ray and UV emission from accretion onto the planet should ionize a significant fraction of the incoming stellar wind, thereby allowing radio waves to be generated from this interaction. 

In reality, the temperature and luminosity based on the Bondi--Hoyle accretion (equations \ref{eq:Lacc} and \ref{eq:Tacc}) describes the situation only approximately.
Precisely, only the neutral portion of the stellar wind can accrete onto the planet without interacting with the planetary magnetic field, and the ionized portion would lose some energy at the bow shock before it accretes.
On the other hand, the ionized plasma interacting with the magnetic field could have a cross-section larger than the Bondi--Hoyle accretion radius.
The detailed electromagnetic structure around RGHJs therefore requires elaborate numerical simulations that are beyond the scope of this paper.
In the following sections, we aim to give order-of-magnitude estimates of radio emission, observability, etc., which ought to be robust with respect to uncertainties in the details of the ionization process.

\section{Estimates}
\label{s:result}

\subsection{Planetary Magnetic Field and Frequency of Radio Emission}
\label{ss:Bplanet}

\begin{figure}[bhtp]
   \plotoneh{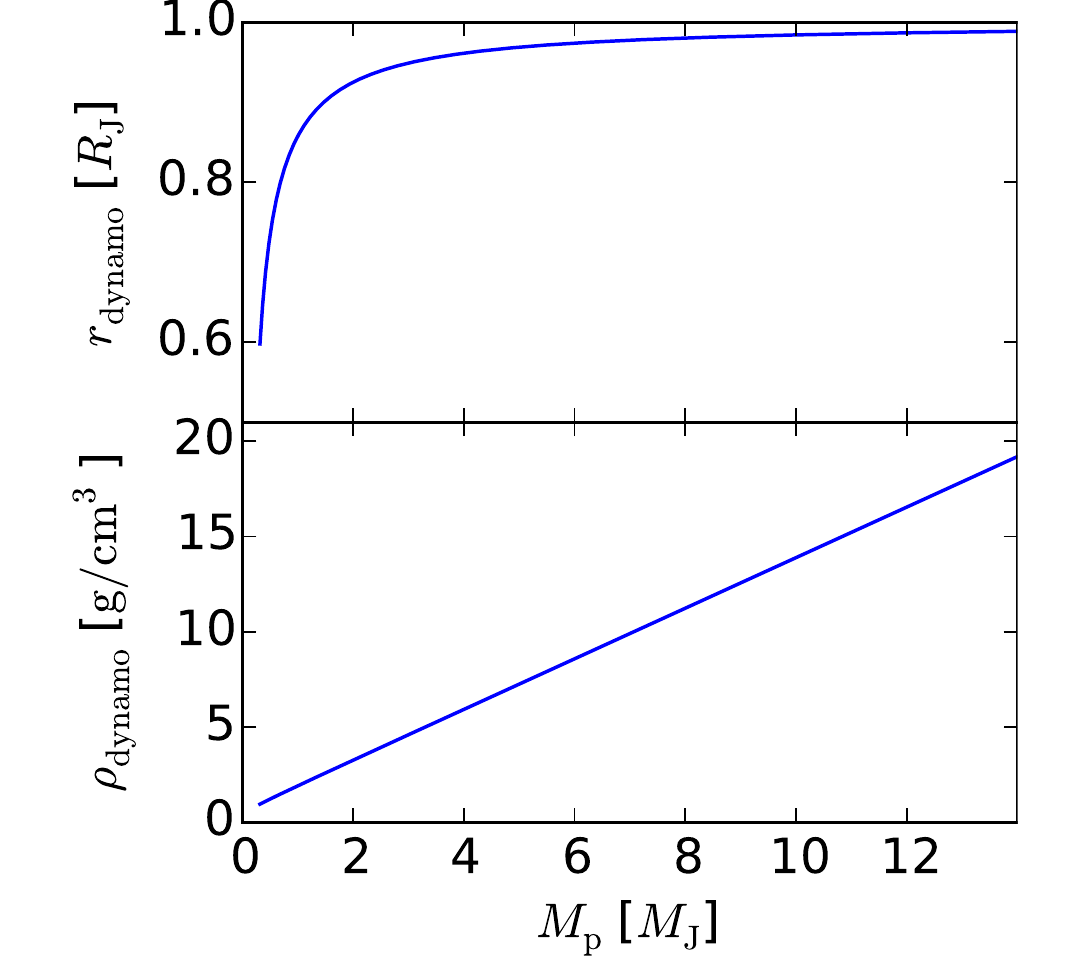}
   \plotoneh{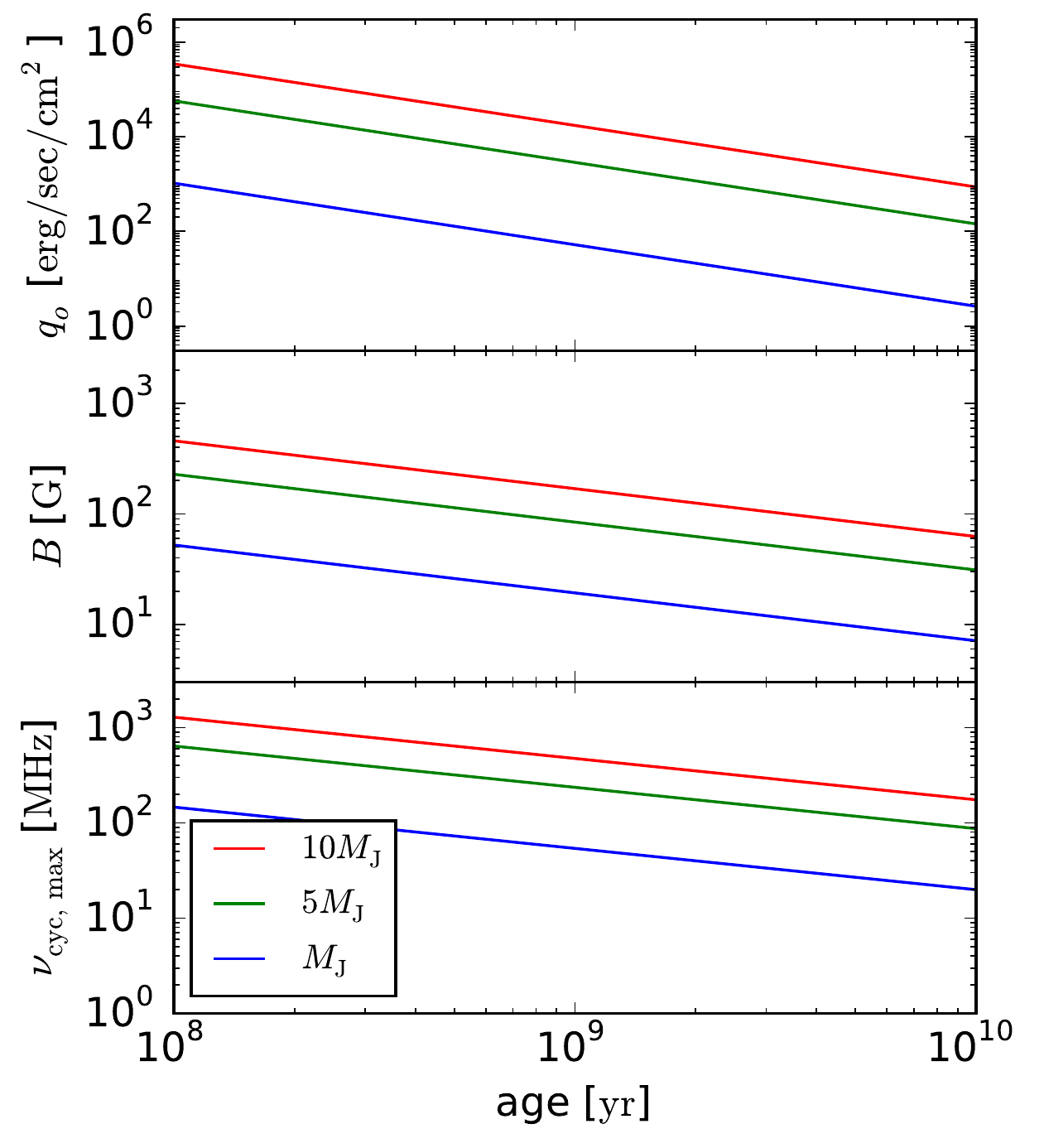}
   \caption{Upper panels: radius and average density of the dynamo region as a function of planetary mass.
     Lower panels: evolutions of heat flux at the outer boundary, planetary magnetic field, and maximum cyclotron frequency, for planets with varying masses ($M_{\rm J}$: blue, $5M_{\rm J}$: green, and $10M_{\rm J}$: red).} 
  \label{fig:planetaryB}
\end{figure}

Figure \ref{fig:planetaryB} shows the computed radius ($r_c$) and average density ($\rho_c$) of the dynamo region as a function of planetary mass, as well as the heat flux at the outer boundary of the core ($q_o$), the estimated strength of the planetary magnetic field ($B$), and the corresponding cyclotron frequency ($\nu_{\rm cyc}$) as functions of planetary mass and the age.
Substituting equation (\ref{eq:burrowsHeatFlux}) into equation (\ref{eq:Bscaling2}), and given that $r_{\rm dynamo} $ does not change significantly, the magnetic field is approximately:
\begin{equation}
B   \sim   B_{\rm J} \left( \frac{M_p }{M_{\rm J}} \right)^{1.04} \left( \frac{t}{4.5~\rm{Gyr}} \right)^{-0.43} \label{eq:scalingB}
\end{equation}
under this model, where $B_{\rm J}$ is the magnetic field strength of Jupiter; in this paper, we roughly consider $B_{\rm J} \sim 10~{\rm G}$.  Reasonably, the resultant values agree with \citet{reiners2010}, who adopted the same scaling law for the planetary magnetic field; we show this figure just for completeness. 
Note that the cyclotron frequency of Jovian planets typically falls between 10~MHz and 1~GHz. 
In this regime, there are a number of current and near-future radio observatories including the Giant Metrewave Radio Telescope (GMRT), Low-Frequency Array (LOFAR), Hydrogen Epoch of Reionization Array (HERA), Square Kilometer Array (SKA), and potential upgrades to the Very Large Array (VLA) (see Section \ref{ss:detectability}).

Since $\nu_{\rm cyc,\,max} > \nu_{\rm plasma}^\oplus $, the radio emission will not be hindered by Earth's ionosphere cut-off. 
On the other hand, it may experience opacity due to the plasma of the stellar wind particles around the planet.
The maximum plasma frequency along the line of sight, $\nu_{\rm plasma}^{\rm los}$, corresponds to that in the vicinity of the planet, if the planet is on the near side of its star to the Earth.
Therefore, substituting equation~(\ref{eq:n_normalized}) to $n_e$ in equation~(\ref{eq:fplasma}), 
\begin{eqnarray}
\label{eq:fplasmalos} \nu_{\rm plasma}^{\rm los} & \sim & 
8979 {\rm~Hz} \times \left( \frac{\dot M_\star}{4 \pi a^2 m_p v_{\rm sw}} \times 1\rm~cm^3 \right)^{1/2} \\
\label{eq:fplasmalos_scaled} &=& 12 {\rm~MHz} \times \left( \frac{a}{5~\mbox{AU}}\right)^{-1} \left(\frac{v_{\rm sw}}{30~\mbox{km/s}}  \right)^{-1/2} \notag \\
 & & \times \left( \frac{\dot M_\star}{10^{-8} M_{\odot }{\rm /yr}}\right)^{1/2} \label{eq:fplasma_RG} \, .
\end{eqnarray}
We can see emission only from where $\nu_{\rm cyc,\,max} > \nu_{\rm plasma}^{\rm los}$. 
The detectable parameter space will be presented in more detail in the next section. 

\subsection{Flux of RGHJ Radio Emission in Comparison with Canonical HJs}
\label{ss:brightness}

\begin{figure*}[bp]
	\plotonesc{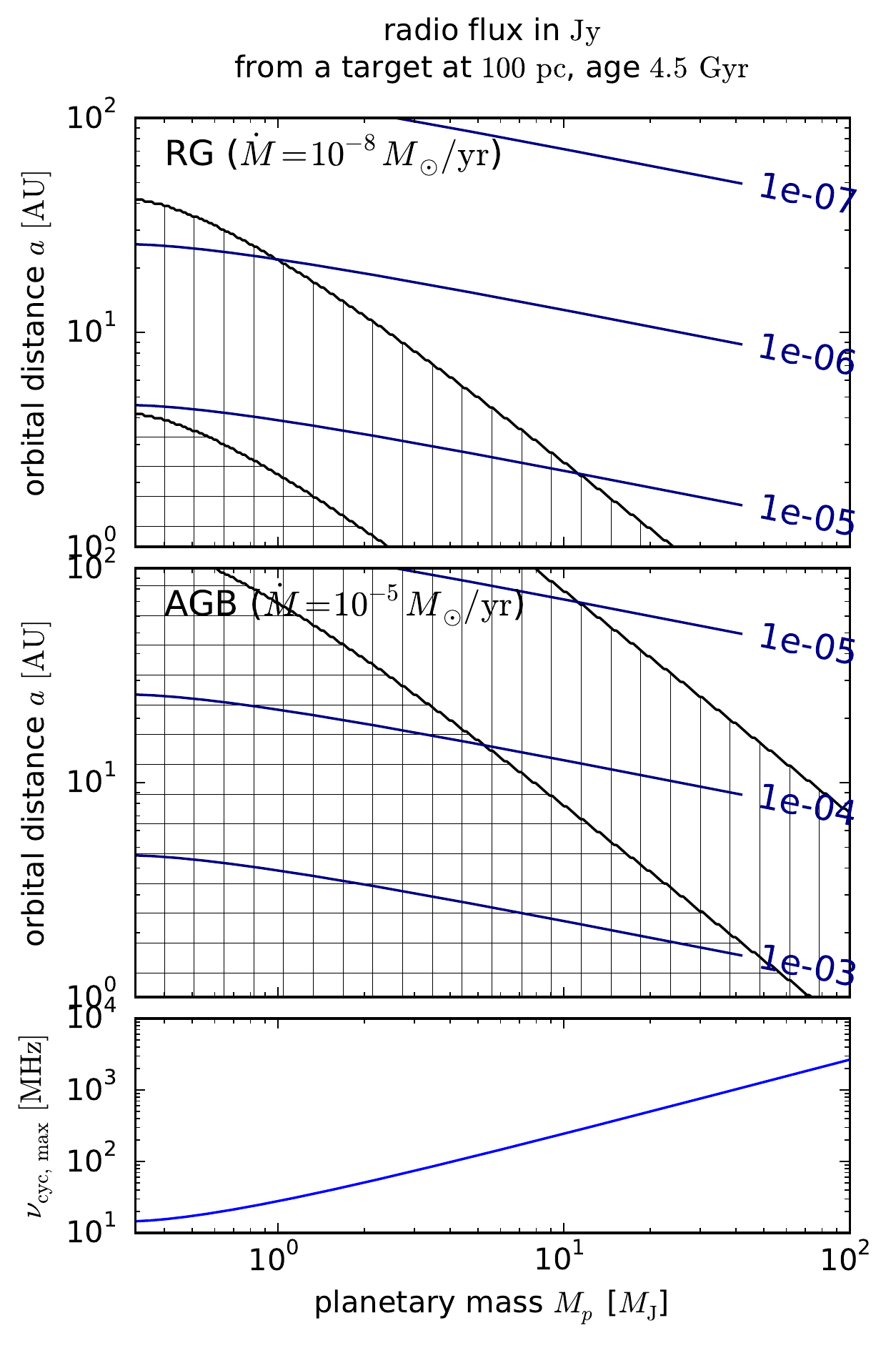}
   \caption{Radio flux in unit of Jy from a planetary companion to a red giant with mass loss rate $10^{-8} M_{\odot }/\mbox{yr}$ (top) and that to an AGB star with mass loss rate $10^{-5} M_{\odot }/\mbox{yr}$ (bottom).
     The systems are located at 100 pc away.
     The doubly-hatched regions show the parameter spaces where the planetary radio emission would not be observable at all because the maximum frequency of the emission (cyclotron frequency at the planetary surface, $\nu_{\rm cyc}$) is below the plasma frequency cut-off, $\nu_{\rm plasma}^{\rm los}$.
     The regions hatched with vertical lines show the parameter spaces where the frequencies of bulk radio emission is below $\nu_{\rm plasma}^{\rm los}$, i.e., $\nu_{\rm plasma}^{\rm los} > 0.1 \nu_{\rm cyc}$.}
  \label{fig:radio}
\end{figure*}

The magnetic stand-off radius (equation \ref{eq:stand-off-radius}) may be written as follows using the stellar mass-loss rate:
\begin{eqnarray}
\nonumber r_{\rm mag} 
&=& r_{\rm mag,\,J} \left( \frac{B}{B_{\rm J}} \right)^{1/3} \left( \frac{a}{5.2\mbox{AU}} \right)^{1/3}  \\
&& \times \left( \frac{\dot M_\star}{\dot M_\odot} \right)^{-1/6}  \left( \frac{v}{v_{\odot }} \right)^{-1/6} \label{eq:stand-off-radius2}
\end{eqnarray}
The typical value for RGHJs is found by substituting relevant values for stellar wind parameters described in Section \ref{ss:stellarwind}:
\begin{eqnarray}
\nonumber r_{\rm mag} &\sim & 14 \, R_J \left( \frac{B}{B_{\rm J}} \right)^{1/3}  \left( \frac{a}{5\mbox{AU}} \right)^{1/3} \\
&& \times \left( \frac{\dot M_\star}{10^{-8} M_{\odot }{\rm /yr}} \right)^{-1/6}  \left( \frac{v}{10^{-1}v_{\odot }} \right)^{-1/6}
 \label{eq:stand-off-radius_RGHJ}
\end{eqnarray}
where we employ $r_{\rm mag,\,J} = 84~R_{\rm J}$ \citep{joy2002}. 

Substituting equation (\ref{eq:stand-off-radius2}) to equation (\ref{eq:Pinp_kin}), the scaling of the radio emission is expanded as follows:
\begin{eqnarray}
P_{\rm k,\,inp} &=& m_{\rm p} nv^3 \cdot \pi r_{\rm mag}^2 \\
&=& P_{\rm k,\,inp,\,J} \left( \frac{B}{B_{\rm J}} \right)^{2/3} \left( \frac{a}{5.2~\mbox{AU}} \right)^{-4/3}  \notag \\
&& \times \left( \frac{\dot M_\star}{\dot M_{\odot}} \right)^{2/3} \left( \frac{v}{v_{\odot}} \right)^{5/3} \label{eq:Pkinp}
\end{eqnarray}

We may compare the radio emission power of RGHJs at 5 AU with that of canonical hot Jupiters set at 0.05 AU.
In order to perform this comparison, equation (\ref{eq:Pkinp}) can be re-normalized as follows:
\begin{eqnarray}
P_{\rm k,\,inp} 
&\approx &  140 ~P_{\rm k,\,inp,\,J} \left( \frac{B}{B_{\rm J}} \right)^{2/3} \left( \frac{a}{5~\mbox{AU}} \right)^{-4/3} \notag \\
&& \times \left( \frac{\dot M_\star}{10^{-8} M_{\odot}{\rm /yr}} \right)^{2/3} \left( \frac{v}{10^{-1} v_{\odot}} \right)^{5/3} \\
&& \;\;\;\;\;\;\;\;\;\;\;\;\;\;\;\;\;\;\;\;\; \mbox{(for RGB stars' companions)} \notag \\
&\approx & 14000 ~P_{\rm k,\,inp,\,J} \left( \frac{B}{B_{\rm J}} \right)^{2/3} \left( \frac{a}{5~\mbox{AU}} \right)^{-4/3} \notag \\
&& \times \left( \frac{\dot M_\star}{10^{-5} M_{\odot}{\rm /yr}} \right)^{2/3} \left( \frac{v}{10^{-1} v_{\odot}} \right)^{5/3}  \\
&& \;\;\;\;\;\;\;\;\;\;\;\;\;\;\;\;\;\;\;\;\; \mbox{(for AGB stars' companions)} \notag \\
&\approx & 490 ~P_{\rm k,\,inp,\,J} \left( \frac{B}{B_{\rm J}} \right)^{2/3} \left( \frac{a}{0.05~\mbox{AU}} \right)^{-4/3} \notag \\
&& \times \left( \frac{\dot M_\star}{\dot M_{\odot}} \right)^{2/3} \left( \frac{v}{v_{\odot}} \right)^{5/3} \label{eq:P_HJ} \\
&& \;\;\;\;\;\;\;\;\;\;\;\;\;\;\;\;\;\;\;\;\; \mbox{(for canonical hot Jupiters)} \notag 
\end{eqnarray}
Here, we have normalized the magnetic field strength of canonical hot Jupiters with $B_{\rm J}$, considering the uncertainty of the magnetic fields of tidally locked planets.
Note that some models of planetary magnetic field strength predict that tidally locked planets have weaker magnetic fields due to their slow rotation \citep[e.g.][]{griesmeier2004}.
Although the orbital velocity of canonical hot Jupiters has been ignored in equation (\ref{eq:P_HJ}), the Keplerian velocity at 0.05~AU around a solar-mass star is $\sim$130~km~s$^{-1}$, which results in only a  $<$10\% increase of the relative velocity.

Using the expressions above, we find that the radio spectral flux density observed at the Earth is:
\begin{eqnarray}
F_{\nu} &=& \frac{P_{\rm radio}}{\Omega d^2 \nu_{\rm cyc}} \\
&\approx & 5.2\times 10^{-8} \mbox{Jy} \left( \frac{d}{100\mbox{~pc}} \right)^{-2}  \notag \\
&&\times \left( \frac{B}{B_{\rm J}} \right)^{-1/3}  \left( \frac{a}{5~\mbox{AU}} \right)^{-4/3} \notag \\
&& \times \left( \frac{\dot M_\star}{\dot M_{\odot}} \right)^{2/3} \left( \frac{v}{v_{\odot}} \right)^{5/3} \label{eq:F_nu} \\
&& \;\;\;\;\;\;\;\;\;\;\;\;\;\;\;\;\;\;\;\;\; \mbox{(for a Jupiter-twin)} \notag \\
&\approx & 0.70 \times 10^{-5} \mbox{Jy} \left( \frac{d}{100\mbox{~pc}} \right)^{-2}  \notag \\
&&\times \left( \frac{B}{B_{\rm J}} \right)^{-1/3} \left( \frac{a}{5~\mbox{AU}} \right)^{-4/3} \notag \\ 
&& \times \left( \frac{\dot M_\star}{10^{-8} M_{\odot}/{\rm yr}} \right)^{2/3} \left( \frac{v}{10^{-1} v_{\odot}} \right)^{5/3} \label{eq:F_nu_RGHJ} \\
&& \;\;\;\;\;\;\;\;\;\;\;\;\;\;\;\;\;\;\;\;\; \mbox{(for RGB stars' companions)} \notag \\
&\approx & 0.70 \times 10^{-3} \mbox{Jy} \left( \frac{d}{100\mbox{~pc}} \right)^{-2}  \notag \\
&&\times \left( \frac{B}{B_{\rm J}} \right)^{-1/3} \left( \frac{a}{5~\mbox{AU}} \right)^{-4/3} \notag \\ 
&& \times \left( \frac{\dot M_\star}{10^{-5} M_{\odot}/{\rm yr}} \right)^{2/3} \left( \frac{v}{10^{-1} v_{\odot}} \right)^{5/3} \label{eq:F_nu_AGB} \\
&& \;\;\;\;\;\;\;\;\;\;\;\;\;\;\;\;\;\;\;\;\; \mbox{(for AGB stars' companions)} \notag \\
&\approx & 2.4 \times 10^{-5} \mbox{Jy} \left( \frac{d}{100\mbox{~pc}} \right)^{-2}  \notag \\
&&\times \left( \frac{B}{B_{\rm J}} \right)^{-1/3} \left( \frac{a}{0.05~\mbox{AU}} \right)^{-4/3} \notag \\ 
&& \times \left( \frac{\dot M_\star}{\dot M_{\odot}} \right)^{2/3} \left( \frac{v}{v_{\odot}} \right)^{5/3} \label{eq:F_nu_RGHJs} \\
&& \;\;\;\;\;\;\;\;\;\;\;\;\;\;\;\;\;\;\;\;\; \mbox{(for canonical hot Jupiters)} \notag 
\end{eqnarray}
Thus, RGHJs are expected to be intrinsically as bright as the closest hot Jupiters. 
Compared with the equivalent Jupiter-like planets around main sequence stars, the massive stellar wind of late red giants can increase the radio emission from planetary companions by more than two orders of magnitude, which allows us to explore 10 times more distant systems, i.e., 1000 times more volume.
This at least partially compensates for the small population of evolved stars. 

Equation (\ref{eq:F_nu_RGHJ}) gives a spectral flux density one to two  orders of magnitude smaller than the prediction of equation (5) in \citet{ignace2010} if we assume the same set of fiducial values and full ionization.
This is primarily because their formulation did not incorporate the effect of a compressed planetary magnetosphere due to a massive stellar wind, while the scaling law for the planetary magnetic field strength is also different.

Using this model for planetary magnetic field strength, we may consider the radio spectral intensity and maximum frequency of a given planetary mass and age. 
Figure \ref{fig:radio} shows contours of radio spectral flux density from planetary companions at 4.5~Gyr with varying masses and orbital distances.
The target system is located at a distance of 100~pc from the Earth; the flux is scaled by the distance as a quadratic function. 

The observable energy flux is limited by the plasma cut-off frequency in the vicinity of the planets, given by equation (\ref{eq:fplasmalos_scaled}).
The doubly hatched regions in Figure \ref{fig:radio} indicate the regions of parameter space where
$\nu_{\rm cyc,\,max} < \nu_{\rm plasma}^{\rm los}$, i.e., the radio emission from the planet cannot be observed from the Earth.  
In reality, the peak of the auroral radio emission does not always occur at the maximum frequency.
Instead, it usually exists at lower frequencies.
Here, we also consider one more criterion, $\nu_{\rm cyc,\,max} > 10 \nu_{\rm plasma}^{\rm los}$ as a conservative measure for the observability of the bulk of radio emission.
In Figure \ref{fig:radio}, this conservative region is shown as hatched regions with vertical lines. 
This indicates that the plasma cut-off due to a dense stellar wind is a significant obstacle to detecting planetary companions around evolved stars.
For a red-giant system, the most promising targets are systems with massive companions with $M_p \gtrsim 5M_{\rm J}$, although systems with smaller companions at distant orbits are also accessible.
For an AGB-star system, however, only very massive planets at distant orbits are marginally detectable, because the stellar wind is so dense that there is significant radio opacity due to the plasma from the accreting stellar wind.

\section{Observability}
\label{s:observability}

In this section, we discuss the observability of the estimated radio emission. 
An obvious potential obstacle is the intrinsic radio emission from host red giant stars themselves; if they are bright relative to the radio flux from the planets, it is significantly more difficult to identify the planetary contribution.
We will see in Section \ref{ss:RGradio} that radio flux from the planets will probably be larger in a certain range of parameter space.
Then, in Section \ref{ss:actualMgiants}, we estimate the radio flux from hypothetical Jupiter-like planets around known M giants. 
In Section \ref{ss:detectability}, we examine the sensitivities and limitations of several current and future
radio instruments at relevant frequencies.
Finally, Section \ref{ss:timevariability} notes the polarization and the time variability of the radio emission as keys to discern the signals from the background noise.

\subsection{Intrinsic Radio Emission of Red-Giants Stars}
\label{ss:RGradio}

\begin{figure}[tbp]
   \plotoneh{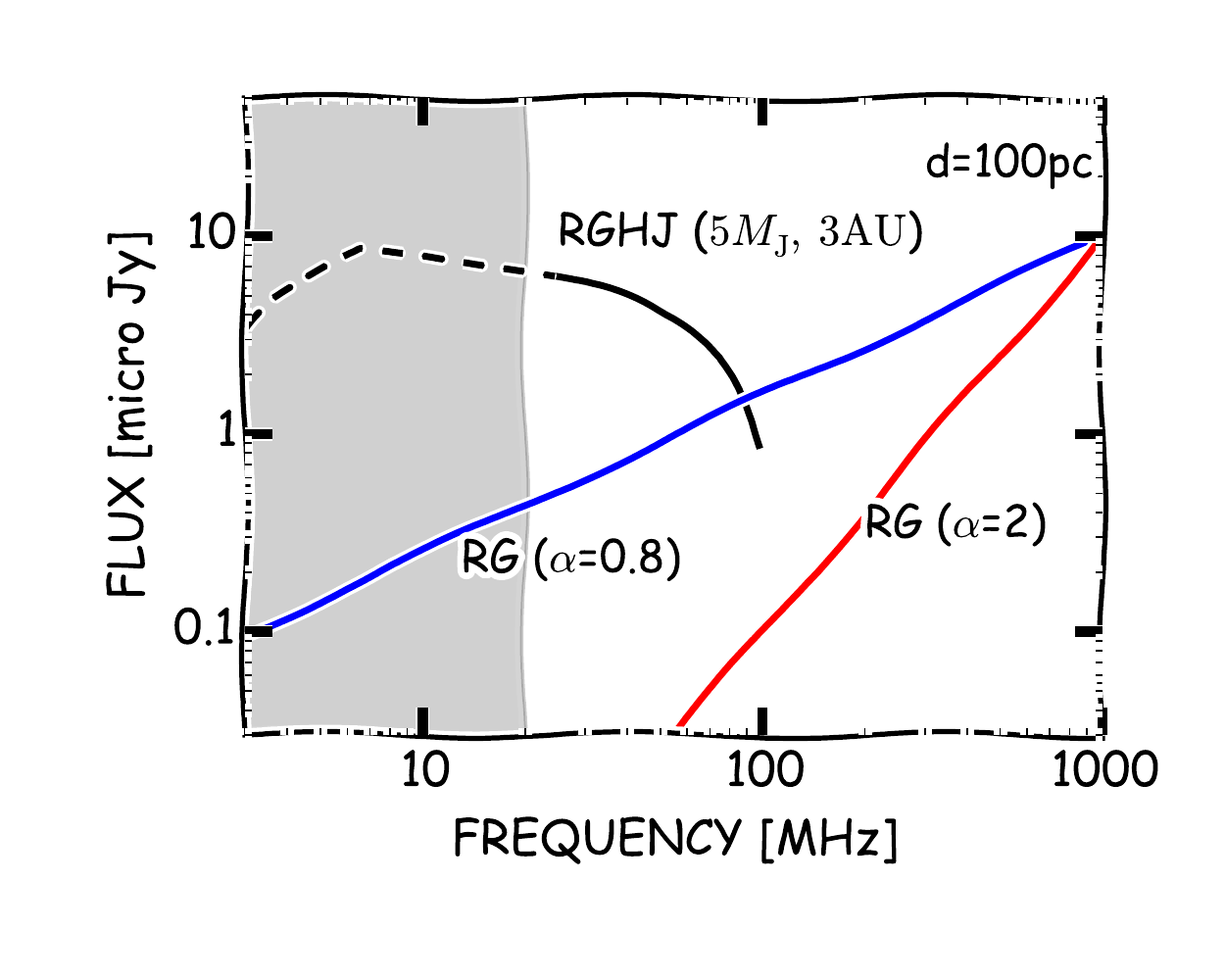}
   \caption{A cartoon of the radio emission spectra of an RGHJ with 5$M_{\rm J}$ and the host red giants with 100 $R_{\odot }$.
     The spectrum of the RGHJ is modeled after Jovian radio spectra, e.g.\ Figure 8 of \citet{zarka_et_al2004} and Figure 2 of \citet{griesmeier2007a}; the contribution from Io-DAM is not shown here.
     The spectra of the host red giant are modeled simply by extrapolating observed radio spectra above 1~GHz with a power law.
     The shaded region represents the portion of the spectrum below the plasma frequency of the stellar wind.}
  \label{fig:cartoon}
\end{figure}

Over the last four decades there have been numerous radio continuum observations of nearby red giants, many undertaken with the aim of understanding the extended atmospheres and mass-loss mechanisms of K- and M-type giants~\citep[e.g.,][]{Newell1982, Knapp1995, Skinner1997, Lim1998, OGorman2013}.
Almost all observations have been carried out at frequencies $\geq$5~GHz, probing thermal Bremsstrahlung emission from the large, partially ionized envelopes surrounding the giant stars~\citep{drake1986}.
The spectral index of this emission is of order unity, with a range of reported values for various sources between 0.8 and 1.6~\citep{OGorman2013}.

Only two published studies describe attempts to detect continuum emission from single (non-binary) red giant stars 
at frequencies below 1 GHz ($L$ band); both yielded null results.
First, a program that observed a sample of nine M-type giants at 430 MHz with the Arecibo Telescope failed to detect any of the sources down to flux density 10 mJy \citep{Fix1976}.
Later, with the Molonglo Observatory Synthesis Telescope (MOST), a sample of eight K- and M-type giants were observed at 843 MHz; the upper limits of their flux densities were placed at approximately 1 mJy \citep{Beasley1992}.
Current facilities could achieve far deeper sensitivity on similar targets.
For example, as listed in Table~\ref{tab:sens}, the GMRT at 150 MHz reaches a sensitivity better than 0.5~mJy in under an hour.
However, any re-attempts to detect single red giants at wavelengths near 1 meter remain to be presented in the literature.

The only red giants that have been detected at meter wavelengths are those in interacting binary systems like those of the RS CVn type.
For example, \citet{vandenOord1994} detected plasma maser emission from II Pegasi at 360 MHz and 609 MHz with the WSRT.
Although single dwarf-type main sequence stars are also known to exhibit bright, coherent radio flares~\citep{Bastian1990}, no analog to this non-thermal emission process has been proposed to exist for evolved stars.

Altogether, we lack firm observational constraints on the brightness of M giants near 30-300 MHz frequencies. 

However, assuming thermal Bremsstrahlung emission continues to dominate in this wavelength regime, we can extrapolate down from the reported centimeter flux densities, using the specific spectral index. 
For example, \citet{OGorman2013} reported $\alpha$~Boo ($R_\star = 25.4 R_{\odot }$, $T=4286$ K, $d=11.26$ pc) at 1~GHz is about 70 $\mu$Jy, and $\alpha$~Tau ($R_\star = 44.2  R_{\odot }$, $T=3910$ K, $d=20.0$ pc) at 3~GHz is about 40 $\mu$Jy. 
Using these data, we extrapolate the radio flux from the host star $F_{\star}$ to the lower frequency range with a simple power law as follows:
\begin{eqnarray}
 F_{\star}(\nu ) & \sim & (0.4-1.4) \times 10^{-5} ~\mbox{Jy} \times \\
\nonumber && \left( \frac{d}{100 ~\mbox{pc}} \right)^{-2} \!\! \left( \frac{R_{\star }}{100~R_{\odot}} \right)^2 \!\! \left( \frac{\nu}{1~\mbox{GHz}} \right)^{\alpha_* } \, ,
\end{eqnarray}
where the power index $\alpha_* $ is of order unity. 

Figure \ref{fig:cartoon} is a cartoon of the radio spectrum of a planet with $5~M_{\rm J}$ and that of the host red giant star with 100~$R_{\odot }$, both of which are placed at a distance of 100 pc.
Note that the thermal contribution from the accretion is negligible in this figure. 
The spectral shape is modeled by simply scaling the Jovian radio spectrum (e.g. Figure 8 of \citealt{zarka_et_al2004}). 
The continuum lines from the star in the cases of $\alpha_* =0.8$ and $2.0$ are shown as two cases.  
As Figure \ref{fig:cartoon} indicates, with a reasonable range of spectral index, the radio emission from the star is smaller by an order of magnitude than the expected radio flux from RGHJs at frequencies below 300~MHz. 
For less massive planets (with weaker magnetic fields), the distinction is even clearer, because the peak flux is larger and the peak frequencies are smaller. 

Therefore, in the following, we ignore the radio emission from the red giant as a noise source.

\subsection{Radio Flux from Nearby M Giants}
\label{ss:actualMgiants}

We now examine how realistic it is that RGHJ radio emission might be detectable with current and near-future instruments.
In order to do so, we consider if known red giant stars were to host planetary companions and estimate the resulting radio flux density.
We use the list of M-type red giants from Table 4 of \citet{dumm1998}, which includes estimates of stellar masses, radii, and effective temperatures.
Their data do not unambiguously distinguish RGB from early AGB stars, so probably about 40\% of the stars on the list are actually early AGB stars.
For each red giant, the mass-loss rate is estimated using the improved Reimers' equation \citep{reimers1975} given by \citet{schroder2005,schroder2007}:
\begin{equation}
\dot M_\star \sim 8 \times 10^{-14} ~M_\odot{\rm /yr} ~\left( \frac{\tilde L \tilde R}{\tilde M}\right) \left( \frac{T_{\rm eff}}{4000} \right)^{3.5} \left( 1 + \frac{g_{\odot }}{4300 g_{\star}} \right) \, , \label{eq:mass-loss}
\end{equation}
where $\tilde L = L_\star/L_\odot$ is the stellar luminosity, $\tilde R = R_\star/R_\odot$ is the stellar radius, $\tilde M = M_\star/M_\odot$ is the stellar mass, and $g_\star$ is the surface gravity. 
The velocity of the stellar wind is assumed to be the escape velocity at $\sim$$4 R_{\star}$:
\begin{equation}
v \sim \sqrt{\frac{2GM_\star}{4R_{\star }}} \, .
\end{equation}
The age of the system is taken to be our crude estimate of the main-sequence lifetime, as a function of stellar mass.
We simply assume:
\begin{equation}
t \sim 10\,{\rm Gyr}\, \left( \frac{M_{\star}}{M_{\odot }} \right)^{-2.5} \, .
\end{equation}
Combining this relation with equations (\ref{eq:fcyc}) and (\ref{eq:scalingB}) results in
\begin{equation}
\nu_{\rm cyc,\,max} \sim 20 \,{\rm MHz}\, \left( \frac{M_{\star}}{M_{\odot }} \right)^{1.075} \left( \frac{M_p}{M_{p,\,{\rm J}}} \right)^{1.04} \, .
\end{equation}
In addition, the distance to the system is obtained based on the parallax data from {\it Hipparcos} data sets\footnote{\url{http://www.rssd.esa.int/index.php?project=HIPPARCOS}}.

Using these data, we estimate the spectral flux of radio emission by specifying planetary mass and orbital distance of a hypothetical planetary companion. 
We consider [A] a Jupiter-twin whose maximum cyclotron frequency is $\nu _{\rm cyc,\,max} \sim 20\,(M_{\star }/M_{\odot})$~MHz, and [B] a larger Jovian planet with $M_p \sim 10M_p$ whose maximum cyclotron frequency is $\nu _{\rm cyc,\,max} \sim 200\,(M_{\star }/M_{\odot})$~MHz.

\begin{figure}[tbhp]
   \plotoneh{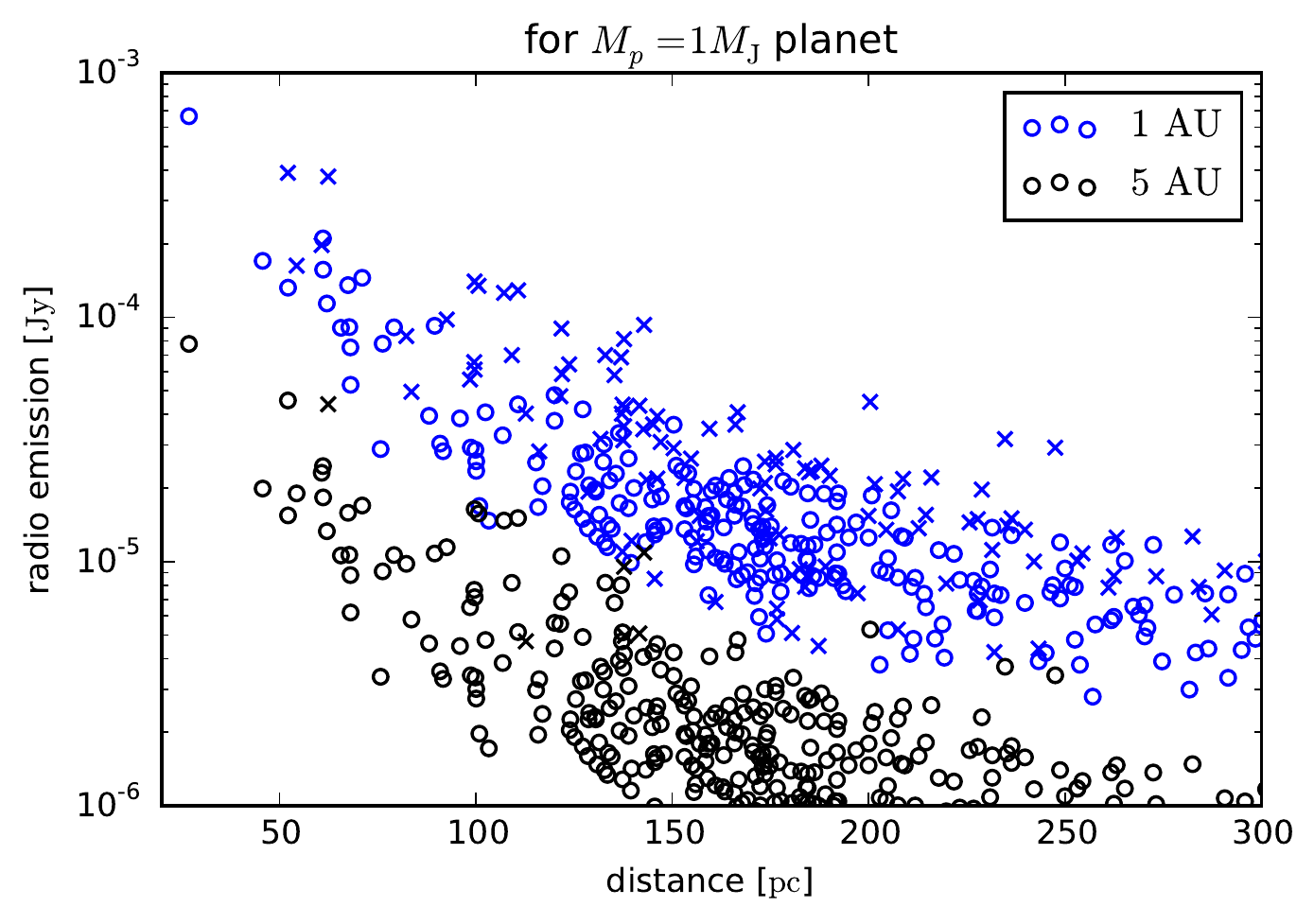}
   \plotoneh{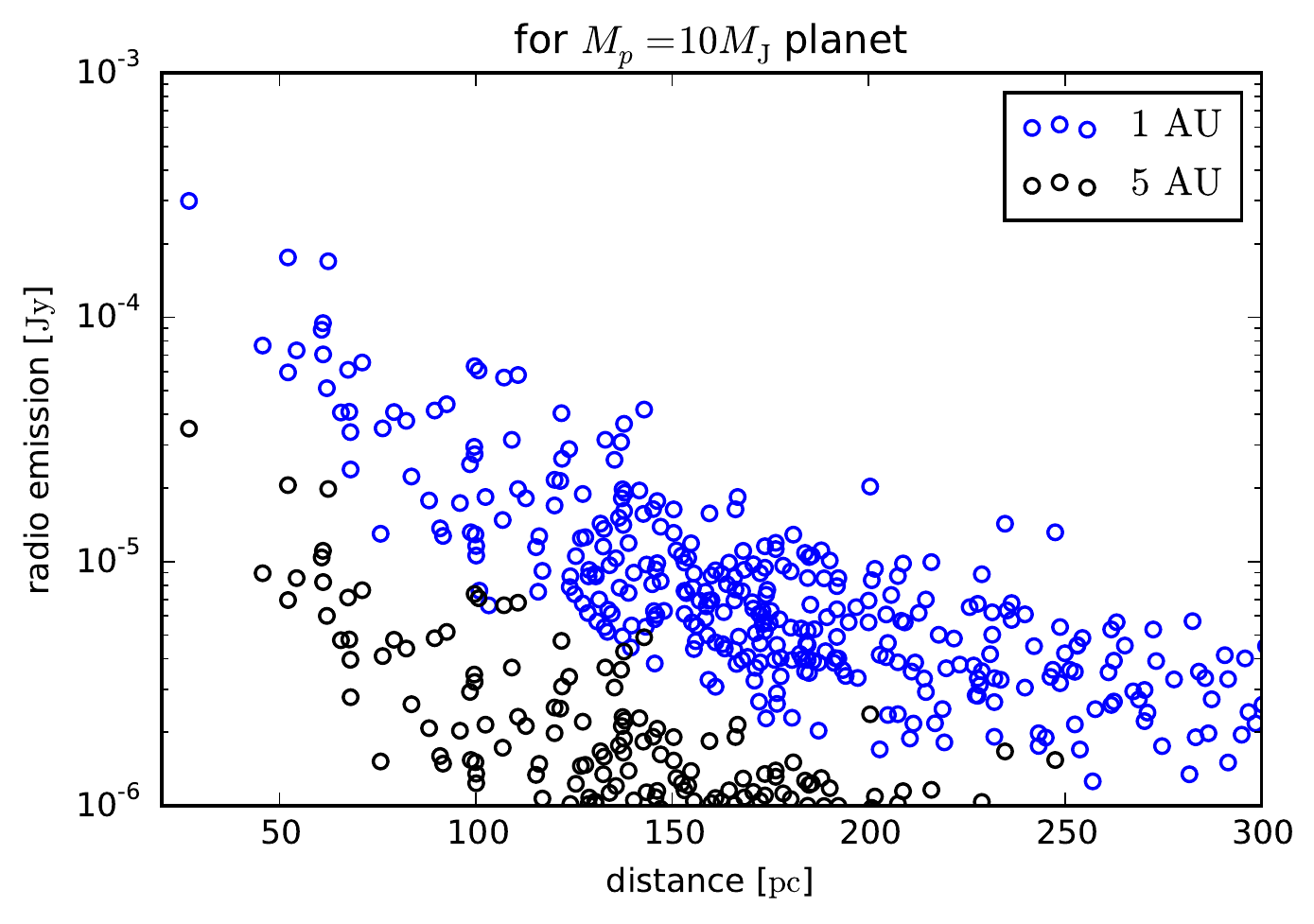}
   \caption{Radio flux from hypothetical RGHJs of 1$M_{\rm J}$ (upper panel) and 10$M_{\rm J}$ (lower panel) at 1~AU (blue) and 5~AU (black).
X symbols indicate that the radio emission is not observable because cyclotron frequency is less than plasma density around the planet. }
  \label{fig:observability}
\end{figure}

Figure \ref{fig:observability} displays the estimated radio flux from planets [A] and [B] around M-type red giants within 300~pc.
To span a reasonable range of orbital distances, we place the planets at 1 AU and at 5 AU from their host stars. 
The shape of the marker indicates whether the radio emission can escape from the system (circles), i.e., $\nu _{\rm cyc,\,max} > \nu_{\rm plasma}^{\rm los}$,  or not (X's).
Assuming Jupiter-mass planets at 1 AU (5 AU) which typically have the maximum cyclotron frequency at 10---100~MHz depending on the age, 148 (12) systems have potential to emit radio waves at a flux density level above $\sim$10$~\mu$Jy out to $\sim$100---300~pc.
Those systems with the highest mass-loss rates, which are intrinsically the brightest systems, would probably be unobservable because of the plasma cut-off frequency of the stellar wind.
For more massive planets, with maximum cyclotron frequency $\sim$100---1000~MHz, the radio flux density is lowered due to its dependence on $B^{-1/3}$ (Equation (\ref{eq:F_nu})), but is more likely to reach Earth because it is less extinguished by plasma opacity.
10-$M_{\rm J}$ planets at $\sim$1~AU may be detectable out to $\sim$200~pc.

A more realistic estimate of the expected number of detections may be evaluated by combining these results with the probability for a star to have a planet with a certain mass and orbit.
Observations of main-sequence stars imply that the number of Jovian planets is in the range of several percent to $\sim$10\% of the number of stars \citep[e.g.][]{cumming2008,cassan2012}.
Assuming the same population for RG systems, the expected number of actual detections from RGHJs would be of order unity. 
A more detailed examination using the empirical planet mass function by \citet{cumming2008} is given in Appendix \ref{s:expectedvalue}.

\subsection{Detectability with Current and Near-future Low-frequency Radio Instrumentation}
\label{ss:detectability}

\begin{deluxetable*}{cccccc}
\tabletypesize{\footnotesize}
\tablecolumns{6}
\tablecaption{Detectability of Hot Jupiters with Current 
and Future Radio Telescopes.\label{tab:sens}}
\tablehead{
Instrument & Band  & \sigRMS\tablenotemark{a} & $\tau$\tablenotemark{b}  & Resolution & $\sigma_c$\tablenotemark{c}\\
           & (MHz) & ($\mu$Jy bm$^{-1}$)         &  (hr)                    & (\arcsec)  & ($\mu$Jy bm$^{-1}$)} 
\startdata
LOFAR-LBA  &  30-80  & 1100  & 230000             & 10   & 57 \\ 
LOFAR-HBA  & 110-200 & 180   &   6500             & 4.1  & 1.1 \\  
HERA       & 50-250  &  76   & 1200               & 410  & 36000\\ 
GMRT       & 130-190 & 162   &   5400             & 16.5 & 57 \\ 
SKA-Low (part) & 50-200 & 5.9 &   7.2             & 7.6  & 9.7 \\
SKA-Low (full) & 50-350 & 1.8 &  0.7              & 4.8  & 1.5 \\
VLA-LOBO   &  50-350 &  77   &  1200              & 8.6  & 10.4
\enddata
\tablenotetext{a}{RMS noise in a 1-hour observation, computed for imaging of the full band reported in this table (Equation~\ref{eq:rms_sens}).}
\tablenotetext{b}{The integration time in hours required to reach an RMS of 2~$\mu$Jy (i.e.\ 5-$\sigma$ detection of a typical source at 100~pc with 10~$\mu$Jy bm$^{-1}$ average flux).}
\tablenotetext{c}{Source confusion RMS contribution.}
\end{deluxetable*}
%
%
%
%

We consider in this section several current and near-future radio observatories that operate at frequencies $\nu \lesssim 350~\rm MHz$ and can reach continuum sensitivities less than 1~mJy in a few hours.
The results are summarized in Table~\ref{tab:sens}. 

The primary limitation to detectability is a telescope's sensitivity, characterized by the root mean square (RMS) noise fluctuations in the sky and receiver.
The RMS noise \sigRMS\ in an interferometric observation with integration time $\tau$, bandwidth $\Delta \nu$, effective area $A_{\rm eff}$ ($\approx0.7 \pi (D/2)^2$ for interferometers comprising antennas with dish size $D$), and $N$ antennas is \citep[see e.g.][]{condon+ransom2016}
\begin{equation}
\sigRMS = \frac{2 k \Tsys}{A_{\rm eff} \, \sqrt{\Delta \nu \, \tau N (N-1)}} 
\left(\frac{\rm 10^{26} Jy}{\rm W m^{-2} Hz^{-1}}\right) \, .
\label{eq:rms_sens}
\end{equation}
Here, \Tsys\ is the blackbody-temperature equivalent of the total system noise, which is the sum of the instrument or receiver noise temperature \Trx\ and the noise contribution from the sky \Tsky.
We estimate \Tsky\ using the low-frequency sky noise temperature fit from \cite{Rogers+Bowman2008}:
\begin{equation}
\Tsky = T_{150} \left( \frac{\nu}{150~\rm MHz}\right)^{-\beta} + \Tcmb
\label{eq:Tsky}
\end{equation}
where $T_{150} = 283.2~\rm K$ is the \Tsky\ at 150~MHz, $\beta=2.47$, and $\Tcmb = 2.73~\rm K$ is the contribution from the cosmic microwave background (CMB).
While this relation was fit to data taken in the 100---200~MHz range, the authors found it to be consistent with published measurements from 10 to 408~MHz. 

An additional factor that affects detectability is source confusion, the imaging noise associated with unresolved interloping radio sources in an observation.
Source confusion is characterized by the standard deviation $\sigma_c$ in the surface brightness of an image due to one or more unresolved sources in the beam solid angle (for a review, see \citealt{Condon1974,Condon2012}).
To estimate $\sigma_c$, reported in Table~\ref{tab:sens}, we use the relations provided by \cite{condon+ransom2016}, reproduced below:
\begin{eqnarray}
\sigma_c & \approx & 0.2 \left( \frac{\nu}{1 \rm~GHz} \right)^{-0.7} \left( \frac{\theta}{1\arcmin} \right)^2\!\!\!\!,~(\theta > 10\arcsec) \\
\sigma_c & \approx & 2.2 \left( \frac{\nu}{1 \rm~GHz} \right)^{-0.7} \left( \frac{\theta}{1\arcmin} \right)^{10/3}\!\!\!\!\!\!\!\!\!\!,~(\theta < 10\arcsec).
\end{eqnarray}
As Table~\ref{tab:sens} illustrates, source confusion is likely to exceed the 2-$\mu$Jy detection limit of a typical RGHJ source for most telescopes under consideration.
However, source confusion can be mitigated through differencing observations that cancel out background sources.
Potential axes for such cancellation are polarization and time.
The source confusion limit is dominated by active galactic nuclei, which have low polarization fractions $\lesssim2.5\%$ \citep{Stil2014} and slowly vary with time.
Therefore, both the circularly polarized nature of the cyclotron emission from planets and the time variability (discussed in \S\ref{ss:timevariability}) partially mitigate the source confusion limit, allowing planetary radio emission, like other time-variable sources, to be detected below the imaging confusion limit.
We therefore expect confusion not to be a strong limitation to the detectability of radio emission from RGHJs.

The LOFAR operates as two separate arrays defined by their high and low bands, differing configurations, and antenna designs, which are called respectively the High Band Antennas (HBA) and Low Band Antennas (LBA).
LOFAR-LBA operates in the 10---90~MHz range, and LOFAR-HBA operates in the 100---240~MHz range \citep{vanHaarlem2013}.
We consider it as two separate instruments here.
LOFAR-LBA's most sensitive band is centered at 60~MHz, and LOFAR-HBA's is at 150~MHz.
Given the strong frequency dependence of the effective aperture of a dipole antenna, and that the receiver noise temperature for LOFAR $\Trx \sim \Tsky$, we use the LOFAR Image Noise calculator\footnote{Heald; \url{http://www.astron.nl/~heald/test/sens.php}}, which is based on {\it SKA Memo 113}\footnote{Nijboer, Pandey-Pommier, \& de Bruyn; \url{http://www.skatelescope.org/uploaded/59513\_113\_Memo\_Nijboer.pdf}}, to obtain the sensitivity and imaging parameters in Table~\ref{tab:sens}.

The HERA\footnote{\url{http://reionization.org/}} is a telescope array under construction in South Africa that will ultimately consist of 352 14-m parabolic dishes, operating from 50 to 250~MHz \citep{pober_et_al2014}.
Unlike the other telescope observatories considered here, HERA is a dedicated experiment for making power spectral measurements of cosmic reionization over wide fields of view.
Similar to its progenitor, the Precision Array to Probe the Epoch of Reionization (PAPER; \citealt{parsons_et_al2010}), HERA's dishes are deployed in an extremely compact configuration \citep{parsons_et_al2012} and statically point toward zenith.
While these features make HERA less ideal for targeted observations, HERA's substantial collecting area, long observing campaigns, and wide field of view make it a powerful survey instrument.

The GMRT is an array comprising 30 25-meter antennas, operating at frequencies 130---2000~MHz.
The lowest band (130---190~MHz) exhibits receiver noise $T_{\rm rx}$ comparable to the sky noise.
To be conservative, we use the GMRT online calculator.\footnote{\url{http://gmrt.ncra.tifr.res.in/~astrosupp/obs\_setup/sensitivity.html}}

The low-frequency component of the SKA, recently rebaselined to have 0.4~km$^2$ collecting area in 130,000 elements with baselines extending to 65~km, will operate from 50 to 350~MHz.\footnote{\url{https://www.skatelescope.org/news/worlds-largest-radio-telescope-near-construction/}}
We estimate the noise in the lower half of the band ($\nu<200~\rm MHz$), appropriate for lower mass ($\lesssim 10 M_{\rm J}$) separately from the full band, which could be used to detect $\sim$30-$M_{\rm J}$ objects.
We again adopt the \cite{Rogers+Bowman2008} relation for \Tsys\ (Equation~(\ref{eq:Tsky})) and assume $\Trx \approx 60~\rm K$.
We caution that these assumptions are optimistic, and the time estimates could in reality be a factor of a few worse.

The low-frequency receivers (28---80~MHz; 4-Band) in the VLA have $\Trx \sim 260 \rm~K$, similar to the receiver noise temperatures for the Long Wavelength Array (LWA; \citealt{Hicks2012}).
As noted in \citet{Hicks2012}, this noise level is subdominant to \Tsky\ by at least -6~dB.
Future upgrades to the VLA such as the LOw Band Observatory (LOBO; \citealt{Kassim2015IAU}) are being considered, and could potentially cover the full 30---470~MHz band.
We estimate the sensitivity of LOBO assuming the receiver noise extrapolates linearly from $\Trx \approx 60~\rm K$ (internal VLA memo) in P-Band (230---470~MHz) to $\Trx \approx 260~\rm K$ in 4-band, and consider just the 50---350~MHz portion of the full 30---470~MHz band.

The noise estimates shown in Table \ref{tab:sens} indicate that, at the frequencies relevant for the detection of radio emission from Jupiter-size planets, current instrument sensitivities are generally too low.
On the other hand, for more massive exoplanets ($M_p \sim 10 M_{\rm J}$) where the bulk of the radio emission would exist above 50~MHz, the low-frequency component of SKA will be able to detect the signal within half a day from planets with $M_p \sim 10~M_{\rm J}$ at $\sim$1~AU within $\sim$200~pc, or those at $\sim$5~AU within $\sim$100~pc. 
VLA-LOBO could also work for a few nearby systems.

\subsection{Discriminating the Signal from the Background}
\label{ss:timevariability}

Since the confusion limit is on the same order as the signal (Table \ref{tab:sens}), it is crucial to consider the ways to discriminate the signal from these sources. 
There are two key features that makes planetary radio emission distinct: the circular polarization and the time variability \citep{hess2011,zarka2015}.  

Planetary auroral radio emission is nearly completely circularly polarized (e.g., \citealt{dessler1983}, and references therein).
In contrast, active galactic nuclei --- the major confusion sources --- have low polarization fractions $\lesssim$2.5\% \citep{Stil2014}.

The other key feature is the significant time variability of the planetary radio emission.
The factors that influence time variability are listed below:

\paragraph{Modulation due to planetary spin rotation ( $\sim$a few hours)}
Radio bursts from Jupiter tend to be correlated with its spin phase \citep[e.g.][]{dessler1983} due to the misalignment between the magnetic axis and the spin axis.
Similarly, high-temporal-resolution observations of an RGHJ system could reveal periodicity at the planet's spin frequency.

\paragraph{Modulation due to the presence of satellites ($\sim$a few days)}
Jupiter's radio bursts are affected by the location of its moon Io, as well \citep{dessler1983}, because Io disturbs the surrounding electromagnetic field.
Likewise, if an RGHJ has a moon that emits plasma, then the radio flux from such a system is likely to be modulated by the orbital motion of the moon \citep[see, e.g.,][]{noyola2014}.

\paragraph{Variability of the stellar wind ($\sim$a few months?)}
If the stellar wind passing the planet is variable, any modulation of the plasma density would also lead to time variability of planetary radio emission.
Such variability might be expected to occur on timescales of several months to half a year, as many M giants have semi-regular periodicities on the order of a few hundred days \citep{Kiss:1999aa}.
\paragraph{Modulation due to orbital motion of the planet ($\sim$a few years)}
If the planetary orbit is close to the edge-on configuration, the radio emission is likely to disappear or significantly dim at certain orbital phases, due to the increased plasma cut-off frequency along the path and/or secondary eclipse \citep{lecavelier_et_al2013}.
The plasma cut-off frequency increases at some phases because, when the planet is on the far side of the star from Earth, the path of the radio emission toward Earth comes through the vicinity of the star, where the plasma density may be high enough to be partially or completely opaque to the emission, depending on the ionization fraction of the stellar wind.

\section{Summary and Discussion}
\label{s:conc}

In this paper, we estimate the radio brightness of distant ``hot Jupiters'' around evolved stars (RGHJs). 
Unlike \citet{ignace2010}, we consider that UV/X-ray photons from accretion onto the planets partially ionize the stellar wind in the vicinity of planets, which would otherwise be mostly neutral for a cool star's wind.
This process yields the free electrons that are crucial to producing the radio emission.
Based on such a picture, the dense stellar wind of an RGB or AGB star would interact with the planetary magnetic field and would add power in the form of kinetic energy into the magnetosphere of an RGHJ.
We find that the intrinsic brightness of radio emission from RGHJs could be comparable to or greater than that of canonical hot Jupiters in close-in orbits around main-sequence stars and $>$100 times brighter than distant Jupiter-twins around main-sequence stars.
This implies that they can be searched for 10 times further away or in a volume 1000 times as large.
This can compensate for the rareness of the evolved stars at least partly.
Thus, RGHJs will serve as reasonable targets in future searches for exoplanetary radio emission. 

A major obstacle to observing this radio emission is the plasma cut-off frequency of the (ionized) stellar wind.
Due to the great density of the stellar wind, the cut-off frequency is as high as $\sim$12~MHz for typical red giants and $\sim$400~MHz for typical AGB stars, making planetary-mass companions to AGB stars difficult or impossible to see via their auroral radio emission. 
The most promising targets are massive planetary companions ($M_p \gtrsim 5 M_{\rm J}$) to red giant stars, with magnetic fields stronger than that of Jupiter (and, therefore, higher cyclotron frequencies). 
Similarly, if the plasma frequency in the region where the bulk of the radio emission is generated is large relative to the cyclotron frequency, this could pose a serious obstacle to observability.

The radio flux from a system with a Jovian exoplanet at a distance of 100~pc would typically be on the order of $\sim$10~$\mu$Jy.
Such signals would be detectable with SKA within half a day. 
For a few nearby systems, a possible upgrade to the VLA such as LOBO would also provide reasonable integration times and resolving power.
In both cases, it is critical to consider polarization and/or time variability of the planetary radio flux, in order to separate it from confused sources in the same resolution element.

The radio emission from RGHJs may soon prove to be a valuable tool for surveying exoplanets in a region of parameter space around highly evolved stars where traditional exoplanet-discovery methods, such as transit and radial-velocity observations, are less sensitive.
Once this planetary radio emission is detected, it will provide a unique approach for studying RGHJ properties.
As the auroral radio flux is significantly modulated in association with several physical processes described in Section \ref{ss:timevariability}, the time variability is useful to infer the planetary spin rotation period or the presence of satellites.
If spectra are obtained, the upper cut-off frequency can tell us the magnetic field strength at the planetary surface, which, when combined with constraints on the planetary mass via, e.g., radial-velocity observations, tests the proposed scaling law for the planetary dynamo. 
Additionally, the lower cut-off frequency of the RGHJ emission could probe the stellar wind properties at the planetary orbit. 
And finally, the planetary radio power in such an extreme environment as an  RG's massive stellar wind will provide valuable data points to test the empirical ``radiometric Bode's law.'' 

\vspace{0.5in}

\acknowledgements

{\sc Acknowledgments}

Y.F. is supported from the Grant-in-Aid No. 25887024 by the Japan Society for the Promotion of Science and from an appointment to the NASA Postdoctoral Program at NASA Goddard Institute for Space Studies, administered by Oak Ridge Affiliated Universities. 
D.S.S. gratefully acknowledges support from a fellowship from the AMIAS group.
J.N. acknowledges support from NASA grant HST AR-12146.04-A and NSF grant AST-1102738.
We thank David Hogg for encouraging us to pursue calculations of exoplanetary radio emission.
We thank Greg Novak for helpful conversations.
We thank Jake VanderPlas for developing the XKCD-style plotting package for matplotlib, and acknowledge its use. 
A.R.P. is grateful for support from NSF grants 1352519 and 1440343.
Y.F. greatly acknowledges insightful and helpful discussions with Tomoki Kimura and Hiroki Harakawa. 
The portion of this research for which T.M. is responsible was performed under a National Research Council Research Associateship Award at the Naval Research Laboratory (NRL).
Basic research in radio astronomy at NRL is supported by 6.1 Base funding.

\bibliography{biblio.bib}

\appendix

\section{Planetary outflow}
\label{ap:outflow}

One might anticipate that, when a planet's host star reaches the red giant stage, the planet's atmosphere could start to escape due both to the increased atmospheric temperature (from increased stellar irradiation, which could lead to temperatures $\gtrsim$1000~K) and to the UV/X-ray radiation produced by accretion from stellar wind.
In this section, we briefly estimate these effects on the circumplanetary configuration. 

The conventional escape parameter is computed for RGHJs as the ratio of escape energy to thermal energy, which is the same as the squared ratio of escape speed to thermal speed:
\begin{equation}
\lambda _c = \frac{G M_p m_{\rm p}}{k T_c r_c} = 220 \left( \frac{M_p}{M_{\rm J}} \right) \left( \frac{T_c}{1000 {\rm K}} \right)^{-1} \left( \frac{r_c}{R_{\rm J}} \right)^{-1}
\label{eq:escape_param}
\end{equation}
where $T_c$ and $r_c$ are the radius and temperature at the exobase (the location above which the mean free path of particles is longer than the pressure scale height). 
In this formalism, $\lambda_c \gg 1$ implies Jeans escape, whereas for $\lambda_c$ approaching unity or below, the escape mechanism becomes a hydrodynamic flow.
Eq.~(\ref{eq:escape_param}) suggests that, in most of the cases we consider, the atmospheric loss is in the Jeans regime.
The rate of Jeans escape is 
\begin{eqnarray}
\Phi _{\rm Jeans} [r_c] &=& \frac{1}{\sigma } \sqrt{ \frac{G M_p}{2 R_p^3} } \left( \lambda _c + 1 \right) \sqrt{ \lambda _c } e^{-\lambda _c}
\end{eqnarray}
where $\sigma$ is the cross-section of hydrogen, here taken to be the geometric cross-section associated with the Bohr radius.
The ram pressure at distance $r$ due to such an outflow is
\begin{eqnarray}
P_{\rm ram} [r] & = & m_{\rm p} \Phi _{\rm Jeans} [r_c] \left( \frac{r}{R_p} \right)^{-2} v_{\rm esc} 
\end{eqnarray}
which indicates $2\times 10^{-7} {\rm~dyne~cm^{-2}}$ even when $\lambda _c = 1$. 
In contrast, the ram pressure of the stellar wind (the left side of equation \ref{eq:stand-off}) is $\sim$3$\times$10$^{-5} {\rm~dyne~cm^{-2}}$, or two orders of magnitude greater than the outflow ram pressure.
The ram pressure of the planetary outflow would, therefore, be negligible.

\section{Ionization Cascade}
\label{sec:AppendixA}

For an electron-hydrogen ionizing collision, we estimate the differential cross-section $\sigma$ as a function of recoil energy $d\sigma / dE_{\rm recoil}$.
We approximate the bound electron as a free electron at rest and use the differential cross-section for the scattering of two electrons (M{\o}ller scattering) in the non-relativistic limit:
\begin{equation}
\label{eq:dsig/dOmeg} \frac{d\sigma}{d\Omega_{\rm CM}} \sim \frac{\hbar^2 c^2\alpha^2}{m_e^2 v_{\rm rel}^2 \sin^4 \theta} \, ,
\end{equation}
where $\alpha$ is the fine structure constant, $v_{\rm rel}$ is the relative velocity, $\theta$ is the scattering angle in the center of mass (CM) frame, and $m_e$ is the electron mass.
The recoil energy (that is, the energy transferred to the electron at rest) is related to the scattering angle via
\begin{equation}
  \label{eq:Erec} E_{\rm recoil}= \frac{1}{4} m_e v_{\rm rel}^2 \left( 1 - \cos \theta \right) \, ,
\end{equation}
and hence
\begin{equation}
  \label{eq:dErec/dOmeg} \frac{dE_{\rm recoil}}{d\Omega_{\rm CM}} = -\frac{1}{4} m_e v_{\rm rel}^2 \, .
\end{equation}
We therefore have
\begin{equation}
  \label{eq:dsig/dErec} \frac{d\sigma}{dE_{\rm recoil}} \sim \frac{\hbar^2 c^2\alpha^2}{m_e v_{\rm rel}^2 E_{\rm recoil}^2} \, ,
\end{equation}
which is largest for small recoil energies.
Of course, $E_{\rm recoil}$ must be larger than $E_{\rm Rydberg}$ for this approximation to be meaningful. 

An objection to the above estimate is that the singularity in the $E_{\rm recoil} \rightarrow 0$ limit arises from the long-range nature of Coulomb interactions, which, of course, is absent in the real problem.
Far away, the ionizing electron sees the dipole moment of the neutral atom.
However, for energy transfer of the order of a Rydberg $\hbar c\alpha/2a_0$ we can estimate the transfer momentum and see that it is larger than the inverse of the Bohr radius $a_0$.
This implies that, in this range, the above estimate may actually be reliable.
The transfer momentum is
\begin{equation}
q^2 =\frac{1}{4} m^2 v_{\rm rel}^2 (\sin^2\theta +(1-\cos \theta)^2)= m^2v_{\rm rel}^2 \sin^2 \frac{\theta}{2} \, .
\end{equation}
Comparison with (\ref{eq:Erec}) shows
\begin{equation}
E_{\rm recoil} = \frac{q^2}{2m} \, .
\end{equation}
Now set $E_{\rm recoil} \sim \hbar c \alpha/a_0$ to get
\begin{equation}
q \sim \sqrt{\frac{m\alpha}{a_0}}\sim \frac{\hbar}{a_0}
\end{equation}
where we used $a_0 = \hbar/\alpha m$.

\section{Back-reaction of a plasma flowing into a magnetic field}
\label{ss:offset}

Given the high stellar wind density of evolved stars, one might anticipate that the large quantity of plasma flowing into a region permeated by a magnetic field would, by spiraling around the field lines, generate an opposing magnetic field that partially cancels the intrinsic planetary magnetic field.
To estimate the strength of this effect, consider a uniform flow of particles of charge $e$, mass $m$, and velocity $v$ into a region of magnetic field $B$, spiraling along the initial magnetic field. 
A single charged particle moving perpendicular to the magnetic field moves in a circular orbit of radius 
\begin{equation}
r=\frac{mv_\bot }{eB} \, .
\end{equation}
which creates a magnetic dipole
\begin{equation}
\mu = \frac{e}{2\pi r/v_\bot} \pi r^2 = \frac{1}{2} e r v_\bot = \frac{mv_\bot^2}{2B} \, .
\end{equation}
The volume integral of the canceling magnetic field $B_c$ generated by a single dipole $\mu$ is
\begin{equation}
\int d^3{\boldsymbol r} B_c = \frac{8\pi}{3} \mu \, .
\end{equation}

Hence, for a number density $n$ of these charged particles, the fraction of the canceling field to the background magnetic field is
\begin{equation}
\frac{B_c}{B} = \frac{n(mv_\bot^2/2)}{3 B^2/8\pi} < \frac{\rho_K}{3\rho_B} \, ,
\end{equation}
where $\rho_K$ and $\rho_B$ are the kinetic energy density and the magnetic energy density, respectively. 

Just inside the stand-off point, 
\begin{equation}
m_p n[r_{\rm mag}] v^2 < \frac{B[r_{\rm mag}]^2}{2\pi }
\end{equation}
and thus $\rho_K/\rho _B < 1$. 
When the particle spirals into the planet, the magnetic pressures increases as 
\begin{equation}
\frac{B^2}{8\pi } \propto \frac{1}{r^6} \, .
\end{equation}
On the other hand, the kinetic energy of the moment increases
\begin{equation}
\frac{mv^2}{2} \propto \frac{1}{r}
\end{equation}
Therefore, unless the density increases more drastically than $1/r^5$, the ratio $B_c/B$ is significantly less than unity.

\section{Expected Number of Detections Taking Account of Planet Occurrence Rate}
\label{s:expectedvalue}

In order to estimate the expected number of RGHJs that might be detectable via near-future radio observatories, we make use of the observed occurrence rate (around main-sequence stars) of Jovian planets.
We consider an empirical power-law planet occurrence rate described as
\begin{eqnarray}
f[M_p,P] &\equiv & \frac{d^2N}{d\log M_pd\log P } \\
&=& \mathcal{C} \left( \frac{M_p}{M_{\rm J}} \right)^{\alpha } \left( \frac{P}{1~{\rm day}} \right)^{\beta } \, ,
\end{eqnarray}
where \citet{cumming2008} found $\alpha = -0.31\pm 0.2$, $\beta =0.26 \pm 0.1$, $\mathcal{C} =1.04\times 10^{-3}$ as best-fit values. %
Using this equation, the probability that a given evolved star in Figure \ref{fig:observability} has an RGHJ with mass larger than threshold $M_{p0}$ whose radio emission may be detectable is
\begin{equation}
\mathcal{P }[>M_{p0}] \sim \int_{M_{p0}} d\log M_p\,\int_{P_{\rm min}[M_p]}^{P_{\rm max}[M_p]} d\log P ~f[M_p, P] \, , \label{eq:detect_prob}
\end{equation}
where $P_{\rm min}[M_p]$ is the minimum orbital period to have a cyclotron frequency larger than the ambient plasma frequency, while $P_{\rm max}[M_p]$ is the maximum orbital period, given the planetary mass, to emit detectable radio flux.
We solve for these critical period values by first solving for the corresponding semimajor axis values, and then transforming to period via Kepler's law, $P_{\{{\rm min,max}\}} = 2\pi \sqrt{a^3_{\{{\rm min,max}\}} / GM_{\star } }$.
For the $a_{\rm min}$, combining equations (\ref{eq:fcyc}), (\ref{eq:scalingB}), and (\ref{eq:fplasma_RG}) leads to
\begin{eqnarray}
\nonumber a_{\rm min} [M_p] &\sim & 2.1~{\rm AU} \left( \frac{M_p}{M_{\rm J}} \right)^{-1.04} \left( \frac{t}{4.5~{\rm Gyr}} \right)^{0.43}  \\
&&\times \left( \frac{\dot M_{\star }}{10^{-8}M_{\odot}/{\rm yr}} \right)^{1/2} \left( \frac{v_{\rm sw}}{30~{\rm km/s}} \right)^{-1/2}  \, .
\end{eqnarray}
For $a_{\rm max}$, equations (\ref{eq:scalingB}) and (\ref {eq:F_nu_RGHJ}) indicate
\begin{eqnarray}
\nonumber a_{\rm max} [M_p] &\sim & 3.8~{\rm AU} \left( \frac{F_{\rm min}}{10~\mu {\rm Jy}} \right)^{-3/4} \left( \frac{d}{100~{\rm pc}} \right)^{-3/2} \\
\nonumber && \times \left( \frac{M_p}{M_{\rm J}} \right)^{-1.04/4}  \left( \frac{t}{4.5~{\rm Gyr}} \right)^{0.43/4} \\
&& \times \left( \frac{\dot M_{\star }}{10^{-8}M_{\odot }/{\rm yr}} \right)^{1/2} \left( \frac{v_{\star }}{10^{-1} v_{\odot }} \right)^{5/4} \, .
\end{eqnarray}

\begin{figure}[t!]
   \plotoneShrinkMed{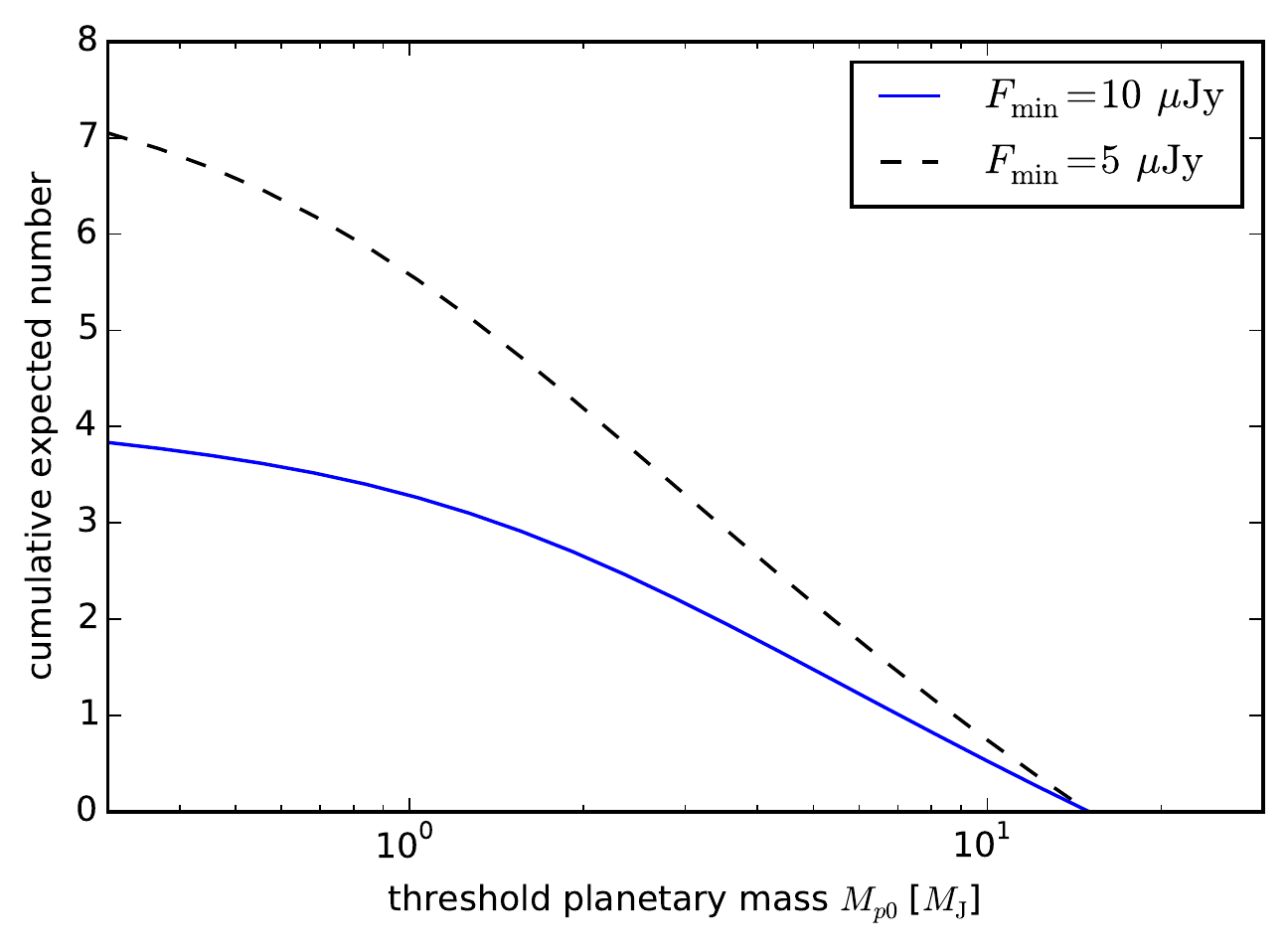}
   \caption{The expected number of radio detection of Jovian exoplanets above the given threshold mass. }
  \label{fig:expected_values}
\end{figure}

Using equation (\ref{eq:detect_prob}), the expected number of planets more massive than $M_{p0}$ with detectable radio emission is $\sum _i \mathcal{P} _i[>M_{p0}]$, where $\mathcal{P}_i[>M_{p0}]$ is the existence and detection probability for the $i$-th target. 
Figure \ref{fig:expected_values} shows the expected number of detections as a function of threshold planetary mass $M_{p0}$.
This figure indicates that, if all of the targets in Figure \ref{fig:observability} are RGB stars, the total number of detections would be about four (seven), provided that the sensitivity of the radio instruments $F_{\rm min}$ is $10~\mu {\rm Jy}$ ($5~\mu {\rm Jy}$). 
However, such high sensitivity of the instrument is unlikely to be achieved at low frequency, so more massive planets are probably the best candidates.
If $\nu _{\rm cyc} > 200~{\rm MHz} $ (i.e., roughly $M_p > 7 M_{\rm J}$) is required for detectability, then the expected number of detectable systems would be reduced to approximately one to two.

\end{document}